\documentclass[a4paper,11pt]{article}
\pdfoutput=1
\usepackage{head,fullpage}     
\usepackage[pdftex]{graphicx}  
\usepackage[utf8]{inputenc}
\usepackage{ulem}
\usepackage{hyperref}
\usepackage{graphicx}
\usepackage{subfigure}
\usepackage{amsmath}
\usepackage{amssymb}
\usepackage{xcolor}
\definecolor{martin}{RGB}{51,153,255}

\definecolor{francesco}{RGB}{0,255,0}

\usepackage{bm}

\begin{document}
\vspace*{1.5cm}
{\noindent\Huge\sffamily\bfseries  Inviscid limit of the   active interface equations}
\vspace*{1cm}

\begin{flushright}
\begin{minipage}[b]{0.85\textwidth}
       {\noindent\Large\sffamily\bfseries Francesco Cagnetta and  Martin R. Evans}\vspace{3pt}
       
       SUPA, School of Physics and Astronomy, University of Edinburgh, Peter Guthrie Tait Road, Edinburgh EH9 3FD, United Kingdom\vspace{3pt}
       
       E-mail: \href{mailto:F.Cagnetta@ed.ac.uk}{F.Cagnetta@ed.ac.uk}\vspace{10pt}
       
       {\noindent\large\sffamily\bfseries Abstract.  } 
We present a detailed solution of the active interface
equations in the inviscid limit.   The active interface equations
were previously introduced as a toy model of membrane-protein systems: they describe a
stochastic interface where growth is stimulated by inclusions which
themselves move on the interface.  In the inviscid limit, the
equations reduce to a pair of coupled conservation laws.  After 
discussing how the inviscid limit is obtained, we turn to the
corresponding Riemann problem---the solution of the set of
conservation laws with discontinuous initial condition. In particular,
by considering two physically meaningful initial conditions, a giant
trough and a giant peak in the interface, we elucidate the
generation of shock waves and rarefaction fans in the system. Then, by
combining several Riemann problems, we construct an
oscillating solution of the active interface with periodic boundaries
conditions.  The existence of this oscillating state reflects the
reciprocal coupling between the two conserved quantities in our system.
\end{minipage}
\end{flushright}

\noindent\hrulefill
\tableofcontents                                
\noindent\hrulefill

\newpage	
\footruleheight{0.5pt}
\rhead{\small\sffamily\bfseries Scaling limits of the active interface equations}
\cfoot{}
\rfoot{\thepage}

\section{Introduction}\label{sec:Intro}

The dynamics of kinetically roughened interfaces is by now  a classic problem of statistical mechanics\cite{family1984kinetics,barabasi1995fractal}.
Though the story goes  back as far as the sixties, with the introduction of the Eden model for growing aggregates~\cite{eden1961}, it was not until the eighties that the problem gained significant traction amongst physicists: see, for instance, the work of Edwards and Wilkinson on the Langevin description of growing interfaces~\cite{edwards1982aa}, or that of Witten and Sander on Diffusion-limited aggregates~\cite{witten1983aa}.
What primarily fostered such an interest was the realisation that interfaces of this kind display a scaling behaviour parallel to that of equilibrium phase transitions~\cite{family1985aa}.
It was, in fact, the related principle of universality that guided Kardar, Parisi and Zhang in proposing a stochastic, partial differential equation for the time-dependent profile of a growing interface---the KPZ equation~\cite{kardar1986aa}.
A moving interface is necessarily out of thermodynamic equilibrium,
nevertheless, the scaling concepts developed for equilibrium problems can be applied with minor, if any, alterations.
The ensuing KPZ universality class, in particular, turned out to include much more than growing interfaces and is now a cornerstone in the physics of driven systems~\cite{kriecherbauer2010pedestrian,halpin2015cocktail}.

A driven system is taken to be held out of equilibrium due to some external driving force.
Many far-from-equilibrium systems, however, especially those inspired by biological problems, are of  a different nature, in that they are kept from relaxing by the continuous energy input at the micro-scales.
They are generically referred to as {\it active} systems, and they have taken centre stage of statistical mechanics in the last few years.
In an endeavour to bridge the theories of active matter and kinetic roughening, we  introduced  in~\cite{cagnetta2018aa} the active interface equations
\begin{equation}\label{eq:ActiveKPZ}
\begin{aligned}
  \partial_t \rho	&=	\quad\Gamma\partial_x\left(\rho \partial_x h\right)	{}&+{}	D\partial^2_{xx}\rho	{}&+{}	\quad\zeta_{\rho},\\
  \partial_t h		&=	\Lambda\rho\left[1-\left(\partial_x h\right)^2\right]	{}&+{}	\nu\partial^2_{xx}h	{}&+{}	\quad\zeta_h,
\end{aligned}
\end{equation}
for the two fields $h(x,t)$ and $\rho(x,t)$, representing the interface height and the inclusion density, respectively ($\zeta_{\rho/h}$ are the corresponding Gaussian noises).
In the specific context of cell membrane dynamics, the height would be the cell's leading edge position, while the density field would refer to proteic inclusions living in the membrane, specifically those that catalyse growth.
We will also adopt the  terminology ``inclusion'' for general considerations beyond this context.

As in the case of the  KPZ equation, which can be obtained from (\ref{eq:ActiveKPZ}) by forcing $\rho(x,t)$ to be constant, the active interface equations can be derived from symmetry considerations.
The new feature is that the $h\leftrightarrow-h$ symmetry breaking, necessary for growth, is effected by the inclusion density rather than an external field.
The physics represented by the active interface equations is fairly simple.
A collection of inclusions (density $\rho(x,t)$) perform overdamped Brownian motion ($D\partial^2_{xx}\rho$ and $\zeta_{\rho}$) within the interface plane.
In addition, each inclusion is coupled to the interface height $h(x,t)$, so that the interface slope is a source of advection for the inclusion density ($\partial_x\left(\rho \partial_x h\right)$).
In fact, the density equation can be derived within the framework of conserved field dynamics~\cite{HohenbergHalperinCriticalDynamics,dean1996aa},
\begin{equation*}
 \partial_t \rho = -\partial_x\left( M(\rho) \partial_x\mu_{\rho}\right) + \partial_x\left(\sqrt{2M(\rho)}\xi\right),
\end{equation*}
where the mobility is given by $M(\rho)=D\rho$ and the chemical potential $\mu_{\rho}$ is related to a free energy $\mathcal{F}(\rho)$,
\begin{equation*}
 \mu_{\rho} = -\frac{d\mathcal{F}}{d\rho},\quad \mathcal{F}(\rho) = \frac{\Gamma}{D}\rho h + \rho \log{\left(\rho\right)},
\end{equation*}
consisting of a linear coupling with the interface height ($\rho h$) and an entropic ideal gas contribution ($\rho\log\rho$).
The interface, in turn, experiences  fluctuations (generated by the  term $\zeta_h$) and a deterministic smoothening due to a surface tension (generated by $\nu\partial^2_{xx}h$).
The effect of the inclusions is that of growth stimulation, and it is represented by the up-down-symmetry-breaking term proportional to $\rho$.
The positive coefficient $\Lambda$ is the signature of the out-of-equilibrium nature of the model: if we try to derive the height equation from the density free energy, even including the surface tension term $\int dx\, \nu\left(\partial_x h\right)^2$, we find a negative $\Lambda$ term.
The non-equilibrium nature of the model is, of course, compatible with the assumption of local energy input: instead of pushing the interface down as free energy minimisation would require, the inclusions exploit the provided energy to lift the interface up.
The consequences of the sign flip are striking: we have explored some of them in~\cite{cagnetta2018aa, cagnetta2019aa}.
The most suggestive emergent behaviour is the organisation of the inclusions into a number of small clusters that generate and surf interface ripples.

In this paper we will focus on the {\it inviscid} limit of the deterministic active interface equations.
The inviscid limit, as the name suggests, neglects ``viscous'' contributions of Laplacian form, that is the diffusion and surface tension in the inclusions/interface language.
This allows us to put dissipation aside for a moment and focus on the deterministic interaction between the fields $h$ and $\rho$.
The inviscid limit  has proved fruitful in the context of interacting particle systems~\cite{GiacominLebowitzPresutti1998}.
For example Burgers' equation---which is the noiseless KPZ equation after a change of variable---has an inviscid limit which reveals phenomena, such as shocks and rarefaction waves, relevant to the full viscous, stochastic equation.
Shocks and rarefaction waves turn out to be fundamental also for the active interface dynamics, though, as we will show, they do not originate from a nonlinearity of the KPZ type.

The construction of solutions to the inviscid active interface equations, obtained by composing shock waves and rarefaction fans, is the central result of this paper.
Namely we find how a wedge-shaped trough is filled in and a wedge-shaped peak is smoothened over.
We then use these solutions on the infinite system to interpret the oscillatory dynamics observed on a finite, periodic system observed in~\cite{cagnetta2018aa}.
Furthermore, we can predict characteristic features such as the interface width oscillation period, the inclusion cluster size and the cluster wave speed, all of which are relevant to possible experiments on real active interfaces.

The remainder of paper is structured as follows. 
In the first part (Section 2), we recall KPZ-like active interface model defined in~\cite{cagnetta2018aa} and describe how the terms in the field equations arise from the microscopic rules.
We close the section by considering the model inviscid limit.
In the remainder of the paper we adopt nonlinear PDEs techniques to solve the inviscid equations.
In particular, we focus on discontinuous initial conditions (the Riemann problem) on the infinite system.
The solutions are presented in detail in Section 3.
We use these solutions to understand the oscillatory dynamics of a finite system in Section 4.
Lastly, in Section 5, we will comment briefly on the effects of noise and viscous terms on the emerging scenario.

\section{The active interface model}\label{sec:model}

\begin{figure}[t!]
	\centering
	\includegraphics[angle=-90,width=1.\textwidth]{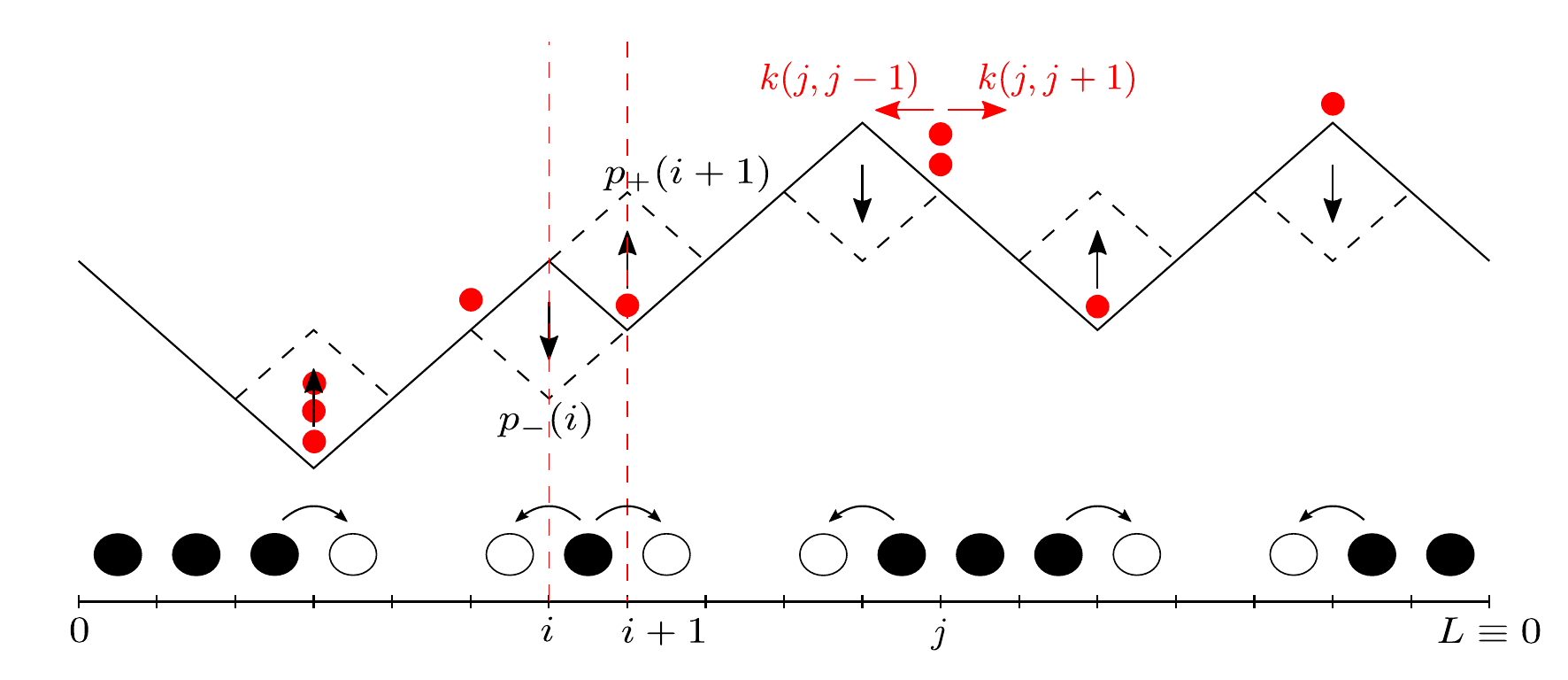}
	\caption{\footnotesize{Schematic illustration of the lattice active interface. Associated with each site there is an height---distance of the black solid line from the reference ticked horizontal line at the bottom---and a number of inclusions---number of red discs sitting on the site. All the possible interface transitions are denoted by black vertical arrows, with a black dashed line representing the outcome of the transition. Particle jumps instead are represented with red horizontal lines. Interface height growth is enhanced by the presence of inclusions and occurs with rate $p_+$, as on the $i+1$-th site in the figure whereas interface height decrease is suppressed by the inclusions and occurs with rate $p_-$, as on the $i$-th 
site  in the figure.  The slopes in the interface advect the inclusions, as on the $j$-th site
where $k(j, j+1) > k(j, j-1)$---the actual rates are defined in the text.}}
	\label{fig:ModelSketch}
\end{figure}
Let us begin with the definition of the microscopic active interface model.
The model was first introduced in~\cite{cagnetta2018aa} and consists of a discrete interface and a collection of inclusions, as shown in Fig.~\ref{fig:ModelSketch} (the interface is the black solid line and the inclusions the red dots).
Both the interface and the inclusions  live on the one-dimensional  lattice, with periodic boundary conditions enforcing the ring topology. The number of lattice sites is $L$ and  we set the lattice spacing to $a=l/L$, so that the ring circumference equals $l$.
The discrete interface is given by a set of heights over the lattice points $\{h_i(t)\}_{i=1,\dots,L}$ and obeys the {\it solid-on-solid} condition~\cite{meakin1986aa,plischke1987aa} $|h_{i+1}(t)-h_i(t)|=1$, which endows the interface with  surface tension.

Each inclusion  is a random walker making jumps  between neighbouring lattice sites.
The walker dynamics is thus specified by  the two (site-dependent) jump rates $k(i,i+1)$ and $k(i,i-1)$, which we define as
\begin{equation}\label{eq:PartRates} k(i,i\pm1) = \left\lbrace\begin{aligned} q_+, &\text{  if  }h_{i}>h_{i\pm1},\\q_-, &\text{  if  }h_{i}<h_{i\pm1}. \end{aligned}\right.\end{equation}
According to Eq.~(\ref{eq:PartRates}), inclusions slide down negative slopes at rate $q_+$ and climb up positive slopes at rate $q_-$.

The interface, in turn,  evolves according to a dynamics which preserves the solid-on-solid condition $|h_{i+1}(t)-h_i(t)|=1$.
According to this condition, each site $i$ of the interface can be a peak, a trough or a slope.
If there is a peak at $i$, then $h_i=h_{i+1}+1=h_{i-1}+1$, so that the interface can be symbolically represented as $\wedge$.
If, instead, $i$ is a trough, then $h_i=h_{i+1}-1=h_{i-1}-1$ and the height profile looks like $\vee$.
Slopes, finally, can be either positive ($h_{i+1}-1=h_i=h_{i-1}+1$, $\diagup$) or negative ($h_{i+1}+1=h_i=h_{i-1}-1$, $\diagdown$).
Each trough can grow and become a peak at rate $p_+(i)$ whereas peaks become troughs at rate $p_-(i)$, so that the solid-on-solid condition is preserved at all times.
To account for the growth-stimulating action of the inclusions, we take the
 interface  rates $p_{\pm}$ to depend on the number of inclusions on the $i$-th lattice site at time $t$, $n_i(t)$ (there is no exclusion interaction between inclusions), i.e. 
\begin{equation}
p_{\pm}(i) = p_{\pm}(n_i)\;.
\end{equation}

A mean-field style derivation of the field equations from microscopic rules analogous to those defined here was given in the supplementary material of~\cite{cagnetta2018aa}. Here we will improve upon that derivation, by considering values of $p_{\pm}$ and $q_{\pm}$ that lead to a factorised steady state: we prove the existence of such steady state in section~\ref{ssec:SS}, then use it to derive the active interface equations in section~\ref{ssec:CG}.

\subsection{Factorised steady state}\label{ssec:SS}

In this section we show that, under a particular condition on the rates, the probability measure
\begin{equation}\label{eq:SSProb}
 P_{ss}(\left\lbrace h_i \right\rbrace,\left\lbrace n_i \right\rbrace) = Z^{-1}_{L,N}\prod_{j=1}^L f(n_j),
\end{equation}
is invariant for the dynamics described above. 
Equation (\ref{eq:SSProb}) states that the stationary state probability is independent of the interface configuration, $\{ h_i\}$ and depends
on the configuration of particles  $\{ n_i\}$ in a factorised way, which implies vanishing  of correlations in the large system limit.
In Eq.~(\ref{eq:SSProb}), $f(n)= \left(q_+ +q_-\right)^{-n}/n!$ and
\begin{equation}\label{eq:Norm}
Z_{L,N}=\sum'_{\left\lbrace h_i \right\rbrace}\sum_{\left\lbrace n_i \right\rbrace}  \prod_{j=1}^L f(n_j)\delta\left(\sum_{j=1}^L n_j-N\right),
\end{equation}
where the $'$ on the $h_i$'s sum represents the solid-on-solid condition $|h_{i+1}(t)-h_i(t)|=1$.
The proof, needed for the following section, follows the lines of the calculation of the steady-state probability of zero-range processes (ZRP)~\cite{evans2005aa}.

In order to determine when Eq.~(\ref{eq:SSProb}) holds,
we first write the master equation for $P_{t}(\left\lbrace h_i \right\rbrace,\left\lbrace n_i \right\rbrace)$ as
\begin{equation}
 \partial_t P_{t}(\left\lbrace h_i \right\rbrace,\left\lbrace n_i \right\rbrace) = \sum_{j=1}^L \left(\mbox{IN}-\mbox{OUT}\right)_j,
\end{equation}
where $\mbox{IN}_j$ and $\mbox{OUT}_j$ are gain and loss terms relative to transition occurring at the $j$-th site, e.g. a change of $h_j$ or an inclusion jumping in or out the $j$-th lattice site.
The interface configuration is completely specified once the locations of local maxima and minima are given (together with the absolute height of one of them). Between a minimum at $i$ and a maximum at $i+k$, for instance, there will be a ``cluster'' of positive slopes. In such a cluster, if $i<j<i+k$,
\begin{equation}
 \begin{aligned} \left(\mbox{IN}-\mbox{OUT}\right)_j = & (n_{j-1}+1)q_- P_t\left(\left\lbrace h_i \right\rbrace,\dots,n_{j-1}+1,n_j-1,\dots\right) \\+& (n_{j+1}+1)q_+ P_t\left(\left\lbrace h_i \right\rbrace,\dots,n_{j}-1,n_{j+1}+1,\dots\right)\\+&\left[n_{j}(q_+ + q_-)\right]P_t\left(\left\lbrace h_i \right\rbrace,\dots,n_{j-1},n_{j},n_{j+1},\dots\right), \end{aligned}
\label{inout}
\end{equation}
The first term in (\ref{inout}) stems from the transition $(n_{j-1}+1,n_j-1)\rightarrow(n_{j-1},n_j)$  (total rate $(n_{j-1}+1)q_-$ on a positive slope), the second from  the transition $(n_{j}-1,n_{j+1}+1)\rightarrow(n_{j},n_{j+1})$ (total rate $ (n_{j+1}+1)q_+$) and the last from $(n_{j},n_{j+1})\rightarrow(n_{j}-1,n_{j+1}+1)$ and $(n_{j-1},n_{j})\rightarrow(n_{j-1}+1,n_{j}-1)$.
As shown in~\cite{evans2005aa}, the contribution to the master equation right-hand side vanish on $P_{ss}$ given by Eq.~(\ref{eq:SSProb}), with
\begin{equation}\label{eq:SSfactors}
 f(n) = \prod_{l=1}^n \frac{1}{l(q_++q_-)} = \frac{\left(q_+ +q_-\right)}{n!}^{-n}.
\end{equation}
Similarly  the same $f(n)$ as above causes $(\mbox{IN}-\mbox{OUT})_j$ to vanish also if $j$ belongs to a cluster of negative slopes, i.e. there is a height maximum at $i$, a minimum at $i+k$ and $i<j<i+k$. 

In the case of maxima and minima of the height, i.e. the aforementioned clusters boundaries, one has  interface transitions at $j$ in addition to the
inclusion transitions.
 If, for instance, $j$ is a height minimum (such as the $i+1$-th lattice site in Fig.~\ref{fig:ModelSketch}), inclusions move from $j\pm 1$ to $j$ at rate $q_+$ while jumping out of $j$ at rate $q_+$, so that the contribution from inclusion transitions is 
\begin{equation}
 \left(\mbox{IN}-\mbox{OUT}\right)^{\rm INCL}_j = 2n_j\left(q_+-q_-\right)Z^{-1}_{L,N}\left[\prod_{i=1}^L f(n_i)\right].
\label{incl}
\end{equation}
The  additional contribution to the master equation, coming from the interface transition (in at rate $p_-$, out at rate $p_+$, cf. Fig.~\ref{fig:ModelSketch}), is simply
\begin{equation}
 \left(\mbox{IN}-\mbox{OUT}\right)^{\rm INT}_j = \left[p_-(n_j) -p_+(n_j)\right]Z^{-1}_{L,N}\left[\prod_{i=1}^L f(n_i)\right]\;,
\label{int}
\end{equation}
because (\ref{eq:SSProb}) does not depend on the interface configuration.
The inclusions and interface terms (\ref{incl}) and (\ref{int}) balance each other if and only if
\begin{equation}\label{eq:SScond}
 p_+(n_j)-p_-(n_j) = 2\left(q_+-q_-\right)n_j.
\end{equation}
One can check that condition (\ref{eq:SScond}) also emerges in balancing interface and inclusions transitions at interface height maxima. Therefore, it is the condition on the model rates which guarantees that the probability in (\ref{eq:SSProb}) is invariant for the model dynamics.

\subsection{Field equations \& inviscid limit}\label{ssec:CG}

In this section we derive the systematic part of (\ref{eq:ActiveKPZ}) for the special choice of parameters $\Gamma=\Lambda$, corresponding to a choice of microscopic rates which satisfies condition (\ref{eq:SScond}) of the previous section. We will make use of the well-known mapping between single-step interfaces and exclusion processes~\cite{meakin1986aa}. The mapping, illustrated in Fig.~\ref{fig:ModelSketch}, transforms the interface into a particle system on the half-lattice $i+1/2$, $i=1,\dots,L$, such that a particle is associated to every negative slope and a hole to every positive slope. By calling $\eta_{i+\frac{1}{2}}$ the occupation number of the particle system sites,\begin{equation}\label{eq:mapping}1-2\eta_{i+\frac{1}{2}} = h_{i+1}-h_{i}.\end{equation}The single-step condition ensures $\eta_{i+\frac{1}{2}}=0,1$ and hence implies an exclusion interaction between particles.

By calling $dJ^{i}_{t}$ the net current of particles from $i-1/2$ to $i+1/2$ between $t$ and $t+dt$, one has\begin{equation}\label{eq:CG1} d\eta_{i+\frac{1}{2}}(t) = -dJ^{i+1}_{t} + dJ^{i}_{t} \equiv -\nabla_i dJ^i_t,\end{equation} where $\nabla_i$ is a shorthand for the lattice gradient. Via the mapping, each particle jump corresponds to an interface transition occurring with a given rate, so that, after incorporating exclusion in the rates
\begin{equation}\begin{aligned}
\left\langle dJ^{i}_t\right\rangle = \left\langle p_-(i)\left[\eta_{i+\frac{1}{2}}\left(1-\eta_{i-\frac{1}{2}}\right)\right] - p_+(i)\left[\eta_{i-\frac{1}{2}}\left(1-\eta_{i+\frac{1}{2}}\right)\right]\right\rangle dt\\
 = \left\langle -\frac{p_+(i)+p_-(i)}{2} \nabla_i \eta_{i-\frac{1}{2}} + \frac{p_+(i)-p_-(i)}{2} \left(\eta_{i-\frac{1}{2}} + \eta_{i+\frac{1}{2}} - 2\eta_{i-\frac{1}{2}}\eta_{i+\frac{1}{2}}\right)\right\rangle dt.
\end{aligned}\end{equation}
As in condition (\ref{eq:SScond}), we set $(p_+(n_i)-p_-(n_i))/2$ equal to $(q_+-q_-)n_i$. We can then perform the averages above with the steady-state measure (\ref{eq:SSProb}), so that the average of the $n_i's$ and of the $\eta_{i+\frac{1}{2}}$ (which are determined by the $h_i's$) factorise. Thus, we are left with
\begin{equation}\begin{aligned}
 \frac{d}{dt} \left\langle \eta_{i+\frac{1}{2}}(t)\right\rangle =& \nabla_i\left( \left\langle \frac{p_+(n_i)+p_-(n_i)}{2}\right\rangle \nabla_i \left\langle \eta_{i-\frac{1}{2}}\right\rangle\right)\\
 &-(q_+-q_-)\nabla_i\left( \left\langle n_i\right\rangle \left\langle \eta_{i-\frac{1}{2}} + \eta_{i+\frac{1}{2}} - 2\eta_{i-\frac{1}{2}}\eta_{i+\frac{1}{2}} \right\rangle \right)
\end{aligned}\end{equation}
As the average is performed with the measure (\ref{eq:SSProb}), also $\left\langle \eta_{i-\frac{1}{2}}\eta_{i+\frac{1}{2}}\right\rangle$ can be factorised (in the limit (\ref{eq:ScalingLimit}) to be discussed below). In addition, we will assume that
\begin{equation}
 \left\langle \eta_{i+\frac{1}{2}}(t)\right\rangle = \eta(x,t)|_{x= a\times(i+\frac{1}{2})}, \quad \left\langle n_{i}(t)\right\rangle = \rho(x,t)|_{x= a\times i},
\end{equation}
where $a$ is the lattice spacing and $\eta(x,t),\rho(x,t)$ smooth functions of $x$, and set, without loss of generality, $p_-=p$ with constant $p$. Hence, by keeping only the leading orders in $a$, we get
\begin{equation}\label{eq:CGEndInt}
 \partial_t \eta(x,t) = -a2(q_+-q_-)\partial_x \left[ \rho\eta\left(1-\eta\right)\right] + a^2 p \partial^2_{xx} \eta + a^2(q_+-q_-)\partial_x \left(\rho\partial_x\eta\right)\;.
\end{equation}
We can, analogously, build an equation as (\ref{eq:CG1}) for $dn_i(t)$, then extract an equation for $\rho(x,t)$ which reads
\begin{equation}\label{eq:CGEndIncl}
 \partial_t \rho(x,t) = a\frac{(q_+-q_-)}{2}\partial_x \left[\rho\left(1-2\eta\right)\right] +  a^2\partial_x \left[\frac{(q_++q_-)}{2}\partial_x \rho\right].
\end{equation}

In order to complete the derivation, we must specify a choice of $q_{\pm}$. Here we consider $q_{\pm}=q\left(q \pm a\gamma\right)$, so that, as in~\cite{cagnetta2018aa},  $a\gamma$ measures the strength of the inclusion advection, which vanishes with the lattice spacing in the continuum limit. As a result, the last term in Eq.~(\ref{eq:CGEndInt}) becomes of order $a^3$ while all the others are of order $a^2$. Thus, in the limit
\begin{equation}\label{eq:ScalingLimit}
  a\rightarrow 0\quad\mbox{and}\quad q a^2 \rightarrow D, 
\end{equation}
after performing the change of variables $2\eta = 1 -\partial_x h$,
and setting $\Gamma \equiv \gamma D$, $\nu = p/q$,
 we obtain the deterministic active interface equations
\begin{equation}\label{eq:ActiveKPZDet}
\begin{aligned}
  \partial_t \rho	&=	\quad\Gamma\partial_x\left(\rho \partial_x h\right)	{}&+{}	D\partial^2_{xx}\rho,\\
  \partial_t h		&=	\Gamma\rho\left[1-\left(\partial_x h\right)^2\right]	{}&+{}	\nu\partial^2_{xx}h\;.
\end{aligned}
\end{equation}
In the general $\Lambda\neq \Gamma$ case, the equations above bear the same relationship to the stochastic active interface equations (\ref{eq:ActiveKPZ}) as the Burgers' equation  bears to the KPZ equation.

We have used a choice of model parameters satisfying  (\ref{eq:SScond}), for which the stationary state factorises, to derive equations (\ref{eq:ActiveKPZDet}).
Let us stress that the derivation is not generally exact, as it assumes no correlations between the $n_i$'s and the $h_i's$---such an assumption is not expected  to hold for all the model parameters choices. However, we propose these equations as an approximation for all parameter values.
 The inviscid limit, obtained with a joint $D,\nu\rightarrow 0$ limit while keeping $\Gamma$ and $\Lambda$ fixed, could be probed by looking at the {\it Euler} scale rather than the diffusive one, i.e. by scaling the microscopic model rates $p$ and $q$ with $a$ while performing the continuum limit $a\rightarrow 0$~\cite{GiacominLebowitzPresutti1998}.
That procedure, however, would require a less transparent definition of the microscopic rules.
Furthermore, it would produce viscous terms of vanishingly small intensity, but different from  the simple Laplacians of Eq.~(\ref{eq:ActiveKPZDet}).
Hence, we will take the  inviscid limit by simply taking  $D,\nu\rightarrow 0$
in the equations obtained at the diffusive scale.
This, as shown in~\cite{GiacominLebowitzPresutti1998}, is still a legitimate way of probing the Eulerian behaviour of a system.


\section{Solving the inviscid limit}\label{sec:Inviscid}

Making a simple change of variable to $u=\partial_x h$, the interface slope rather than the bare height, yields the form of the inviscid active interface equations that we consider in this paper,
\begin{equation}\label{eq:PREInviscidActInt}
 \partial_t\begin{pmatrix}\rho\\u\end{pmatrix} + \partial_x\begin{pmatrix}-\Gamma \rho u \\-\Lambda \rho(1-u^2)\end{pmatrix} = 0.
\end{equation}
We now   replace the driving term $\Lambda \rho(1-u^2)$ in the height equation by
$\Lambda \rho$
i.e. we neglect the  KPZ-like nonlinearity $\left(\nabla h\right)^2$
in the height equation. The reason for doing this is that we expect $\Lambda \rho$ to be the leading driving term as it cannot be transformed away by a 
shift of frame of reference (as can  a constant driving term in the usual KPZ equation \cite{kardar1986aa}).
The coupled equations we consider are then
\begin{equation}\label{eq:InviscidActInt}
 \partial_t\begin{pmatrix}\rho\\u\end{pmatrix} + \partial_x\begin{pmatrix}-\Gamma \rho u \\-\Lambda \rho \end{pmatrix} = 0.
\end{equation}
Recalling the elementary fact that a conservation law for a field $\varphi$ in one dimension has the form
\begin{equation}
\label{cons1d}
\partial_t\varphi + \partial_x J_{\varphi} =0,
\end{equation}
we see that (\ref{eq:InviscidActInt})  has the 
the structure of a system of   coupled conservation laws
\begin{equation}\label{eq:InviscidAImat}
 \partial_t\begin{pmatrix}\rho\\u\end{pmatrix} + \partial_x\begin{pmatrix} J_\rho\\J_u\end{pmatrix} = 0,
\end{equation}
where the inclusion current
$J_{\rho} = -\Gamma \rho u $  is proportional to the negative slope and the
$u$ current $J_u = -\Lambda \rho$ is proportional to minus the inclusion density.

\subsection{Summary of solutions of Eq.~(9)}

\begin{figure}[t!]
	\centering
	\includegraphics[angle=-90,width=1.\textwidth]{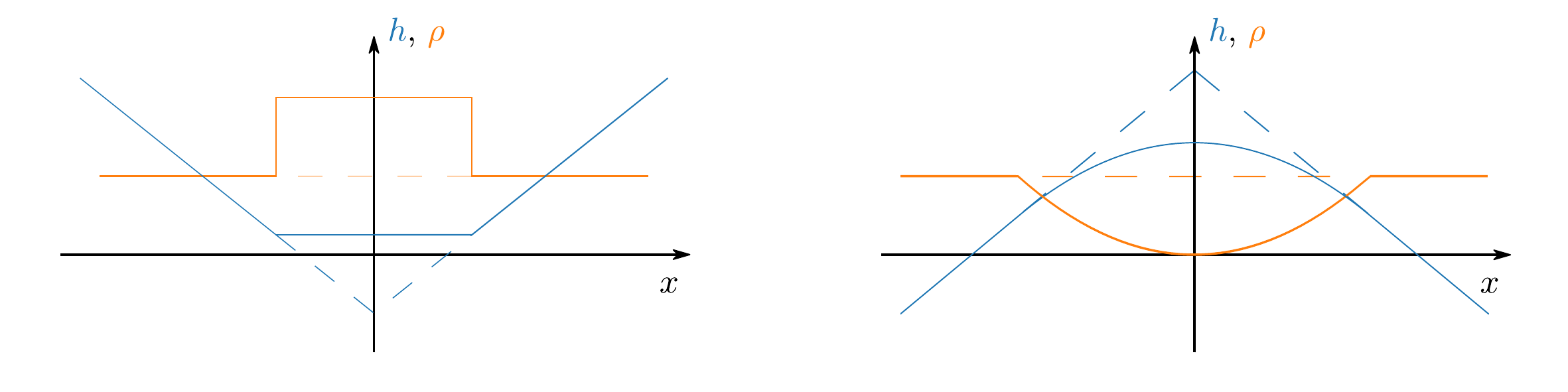}
	\caption{\footnotesize{The anticipated solutions, giant trough (left) and giant peak (right), shown in the height-density variables (solid lines, blue for height and orange for density). Dashed lines denote the corresponding initial condition: while the trough is filled with inclusions as time progresses, the peak is depleted. Note that the trough in the height generates  a pair of shock waves in the density profile whereas the peak in the height generates a pair of rarefaction waves.}}
	\label{fig:AnteSol}
\end{figure}
In the remainder of the section we will solve Eq.~(\ref{eq:InviscidActInt}) for some special initial conditions on the infinite line---let us summarise here the two main cases.
The first is a trough in the interface with a uniform density of inclusions  $\rho(x,t=0)=1$---we call it `giant trough'. The solution reads
\begin{equation}\label{eq:AnteSol1}
 \left(\rho,u\right)(x,t) = \left\lbrace\begin{aligned}&\left(1,-1\right) ,& &x/t\; <-\sqrt{\Lambda\Gamma},\\&\left(1+\sqrt{\frac{\Gamma}{\Lambda}},0\right) ,&-\sqrt{\Lambda\Gamma}\quad <\; &x/t\; < \sqrt{\Lambda\Gamma},\\&\left(1,+1\right) ,&\sqrt{\Lambda\Gamma}
\quad <\; &x/t,\end{aligned}\right.
\end{equation}
and is shown in the left panel of Fig.~\ref{fig:AnteSol} (initial condition dashed, solution solid). According to Eq.~(\ref{eq:AnteSol1}), inclusions accumulate at the bottom of the trough (excess density $\sqrt{\frac{\Gamma}{\Lambda}}$) whilst rendering the height profile flat.
The second initial condition we consider is a peak uniformly filled with inclusions, thus named `giant peak', whose solution, in the simplifying case $\rho(x,t=0)=3\Gamma/4\Lambda$, is given by
\begin{equation}\label{eq:AnteSol2}
 \left(\rho,u\right)(x,t) = \left\lbrace\begin{aligned}&\left(\tfrac{3\Gamma}{4\Lambda},+1\right) ,& &x/t<-\tfrac{3\Gamma}{2},\\&\left(\tfrac{(x/t)^2}{3\Gamma\Lambda},-\tfrac{2}{3\Gamma}(x/t)\right) ,&-\tfrac{3\Gamma}{2}<&x/t< 0,\\&\left(0,0\right) ,& &x=0,\\&\left(\tfrac{(x/t)^2}{3\Gamma\Lambda},-\tfrac{2}{3\Gamma}(x/t)\right) ,& 0<&x/t<\tfrac{3\Gamma}{2},\\&\left(\tfrac{3\Gamma}{4\Lambda},-1\right) ,&\tfrac{3\Gamma}{2}<&x/t.\end{aligned}\right.
\end{equation}
The solution~(\ref{eq:AnteSol2}) is illustrated in the right panel of Fig.~\ref{fig:AnteSol} (initial condition dashed, solution solid) and describes how the inclusions move away from the peak  in the height profile which in turn smoothens.

Although in both these initial conditions and solutions we have $u\sim\mathcal{O}(1)$, in apparent contradiction to the neglect of the $u^2$ term,  $u$ can be made as small as required by rescaling the parameters $\Gamma$ and $\Lambda$. Our approximation then holds as long as $u$ remains bounded at later times, which is true for Eq.~(\ref{eq:AnteSol1}) and Eq.~(\ref{eq:AnteSol2}), and also for the more complex oscillatory solution we build in section \ref{sec:osc}.

The remainder of the section has the following structure.
First, we will review the method of characteristics and how shock waves and rarefaction fans arise in one-dimensional conservation laws.
In section~\ref{ssec:linconlaw} we  define the Riemann problem: solving a {\it system} of conservation laws with a discontinuous initial condition.
Having more than one conservation law complicates the application of the methods of characteristics.
We illustrate some of these complications by considering a simpler model---the linearised active interface equations.
Finally, in section~\ref{ssec:Rp} we study the full inviscid active interface equations by combining concepts from the two previous sections.
We will specifically show that the two discontinuous initial conditions considered, the giant trough and the giant peak, generate shock and rarefaction waves, respectively.

\subsection{Method of characteristics for one-dimensional conservation laws} \label{sssec:1dconlaw}

Conservation laws such as Eq.~(\ref{cons1d}) are convenient mathematically as they can be solved systematically  (at least to obtain an implicit solution)  by applying a simple specialisation of the {\it method of characteristics} which we now outline~\cite{evans2010pde}.
Specifically, conservation laws are quasi-linear equations, i.e. linear in the highest-order derivatives present (here $\partial_t \varphi$ and $\partial_x \varphi$), although the coefficients  may depend on the field $\varphi$ and $x$, $t$.
If,  in conservation laws such as (\ref{cons1d}), $J_\varphi$ depends only on the field $\varphi$, then the characteristic curves  are simply  straight lines $x = J'(\varphi) t+ const.$ and  the solution of the  first order conservation law is constant along such lines.
Then, the method of characteristics reduces to looking at the initial condition at $t=0$ and propagating it at further times on characteristic lines.
However, inconsistencies may appear that prevent regions of the $(t,x)$ plane being filled with characteristics, as, for instance, when some of them cross.

Generally, the crossing of characteristics implies the emergence of discontinuities.
Such discontinuities can propagate in time as {\it shock waves}, provided the conservation law is still satisfied~\cite{evans2010pde}.
This forces the speed $\sigma$ of the propagating discontinuity to satisfy the {\it Rankine-Hugoniot} condition
\begin{equation}\label{eq:1DRankineHugoniot}
 J(\varphi_r) - J(\varphi_l) =\sigma(\varphi_r-\varphi_l),
\end{equation}
where $\varphi_l$ and $\varphi_r$ denote the field values on the left and right of the discontinuity, respectively.
Notice that propagating discontinuities only solve the original conservation law in the  weak sense. That is, they satisfy
\begin{equation}
\int_0^\infty {\rm d}t \int_{-\infty}^\infty {\rm d}x \,
v \left[ \partial_t\varphi + \partial_x J_{\phi}\right]  =0\;,
\end{equation}
for any smooth test function $v(x,t)$ rather than the original equation Eq.~(\ref{cons1d}).
Therefore, discontinuous solutions  might be artefacts rather than actual solutions.
One possible way of identifying the physical solution  is the {\it entropy condition} due to Lax\cite{lax1973hyperbolic},
\begin{equation}\label{eq:1DLaxEntropy}
 J'(\varphi_l) \geq \sigma \geq J'(\varphi_r)
\end{equation}
where $\varphi_{l,r}$ are the densities at  the left and right of the shock respectively.
Eq.~(\ref{eq:1DLaxEntropy}), which can be derived under the assumption of a vanishingly small viscous term (cf. Eq.~(\ref{eq:ActiveKPZDet})), has to be satisfied by the candidate shock.
If not, there will be a {\it rarefaction wave} (sometimes referred to as rarefaction fan): rather than propagating, the discontinuity relaxes through a family of diverging characteristics emanating from the discontinuity.

\subsection{Riemann problem for coupled linear conservation laws}\label{ssec:linconlaw}

In this section we address the coupled equations, Eq.~(\ref{eq:InviscidActInt}), and define the associated Riemann problem.
Let us call $\mathbf{v}$ the vector having $\rho$ and $u$ as components, and $\mathbf{J}(\mathbf{v})$ the corresponding vectorial current.
By defining a matrix $\mathbf{A}({\bf v})$ such that $A_{i,j}=\partial_{v_j}J_i$, Eq.~(\ref{eq:InviscidActInt}) can be written as
\begin{equation}\label{eq:InviscidActIntVecLin}
 \partial_t\mathbf{v} +  \mathbf{A}\cdot \partial_x\mathbf{v} = \mathbf{0}.
\end{equation}
Consider the discontinuous initial condition
\begin{equation*}
 \mathbf{v}(x)=\left\lbrace\begin{aligned} &\mathbf{v}_l, &x<0, \\ &\mathbf{v}_r, &x>0. \end{aligned}\right.
\end{equation*}
The solution of a first order conservation law  equipped with such a step-like initial datum is called a {\it Riemann problem}~\cite{evans2010pde}.
We thus refer to our  step-like initial condition as the $(\mathbf{v}_l,\mathbf{v}_r)$ Riemann problem, with the convention that $\mathbf{v}_l$ is the vector of the system variables on the discontinuity's left and $\mathbf{v}_r$ is the vector on the right.

To be specific, we will consider  a system size-wide wedge-shaped  trough in the interface  with a uniform inclusion density, i.e. $u_l = 1 = -u_r$ and $\rho_l=\rho_r=\rho_0$. We refer to this initial condition as the giant trough (see Figure~\ref{fig:TroughIC}).
As an introductory example let us consider the linearised version of Eq.~(\ref{eq:InviscidActInt})
\begin{equation}\label{eq:InviscidActIntLin}
 \partial_t\begin{pmatrix}\rho\\u\end{pmatrix} + \partial_x\begin{pmatrix}-\Gamma' u \\-\Lambda \rho\end{pmatrix} = 0,
\end{equation}
where $\Gamma'=\Gamma\rho_0$ and $\rho_0$ is the homogeneous density about which we have linearised the equations.
Eq.~(\ref{eq:InviscidActIntLin}) would be relevant in the limit of high density where variations in density are relatively small. For the linearised problem, we have
\begin{equation}
\mathbf{A^{\rm lin}} = \begin{pmatrix} 0 & -\Gamma'\\ -\Lambda & 0 \end{pmatrix}\;.
\end{equation}

\begin{figure}[t!]
	\centering
	\includegraphics[angle=-90,width=1.\textwidth]{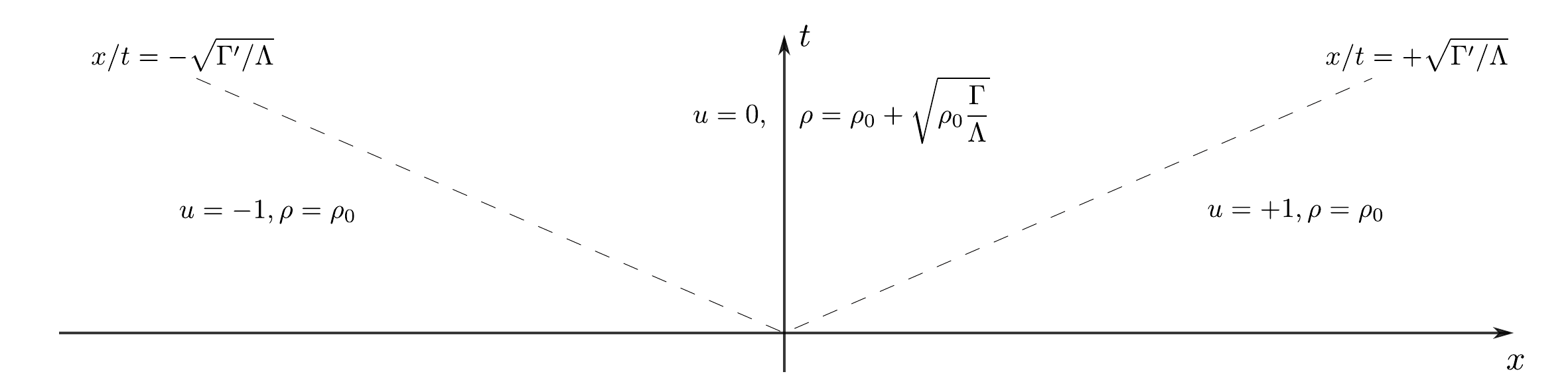}
	\caption{\footnotesize{Solution of the Riemann problem associated with the linearised, inviscid active interface equations. The initial condition is a uniform density ($\rho_l=\rho_r=\rho_0$) wedge ($u_l = 1 = -u_r$, recall $u$ is the interface slope). As time progresses, inclusions accumulate at the bottom of the wedge and make it grow.}}
	\label{fig:InviscidActIntLinSol}
\end{figure}

Now call $\lambda_i$ the $i$-th eigenvalue of $\mathbf{A^{\rm lin}}$ and $\mathbf{l}_i$ the corresponding left eigenvector ($\mathbf{l}_i\cdot\mathbf{A^{\rm lin}}=\lambda_i\mathbf{A^{\rm lin}}$), and assume the eigenvalues to be labelled from the smaller to the larger.
By multiplying from the left Eq.~(\ref{eq:InviscidActIntVecLin}) (and the initial condition) with the left eigenvector, the two-dimensional system decomposes into two independent ones
\begin{equation}\label{eq:LinWaveDecomposition}
 \left\lbrace\begin{aligned}&\partial_t e_i + \lambda_i \partial_x e_i =0, \\ &e_i(x,0)= e^0_i(x),\end{aligned}\right.
\end{equation}
where the $e_i$'s are the system eigenmodes $\mathbf{l}_i\cdot\mathbf{v}$.
The system of equations (\ref{eq:LinWaveDecomposition}) has solutions $e_i(x,t) = e_i^0(x-\lambda_i t)$, which can be composed with $\mathbf{A^{\rm lin}}$ right eigenvectors to yield $\rho(x,t)$ and $u(x,t)$.
To sum up, the Riemann problem is simply solved by projecting the initial discontinuity onto the system left eigenvectors, transporting the projections along $x=\lambda_i t$ and gluing them together again with the right eigenvectors.
The solution for the giant trough initial condition reads
\begin{equation*}
 \mathbf{v}(x,t)=\left\lbrace\begin{aligned} &\mathbf{v}_l, &x<\lambda_1 t, \\ &\mathbf{v}_l + \left[\mathbf{l}_1\cdot(\mathbf{v}_r-\mathbf{v}_l)\right]\mathbf{r}_2, &\lambda_1t<x<\lambda_{2}t,\\ &\mathbf{v}_l +\displaystyle\sum_{i=1,2} \left[\mathbf{l}_i\cdot(\mathbf{v}_r-\mathbf{v}_l)\right]\mathbf{r}_i \equiv \mathbf{v}_r, &\lambda_2t< x\;,\end{aligned}\right.
\end{equation*}
and is shown in Fig.~\ref{fig:InviscidActIntLinSol} on the $(t,x)$ plane.
The fully nonlinear equations will be considered in the next section.

\subsection{Riemann problem for the nonlinear active interface equations}
\label{ssec:Rp}

We have learnt in the previous section that the eigenspaces of of $\mathbf{A}$ play a prominent role in the solution of the linearised Riemann problem.
This also holds in the nonlinear active interface equations, for which
\begin{equation}\label{eq:ActIntMatrix}
\mathbf{A}(\mathbf{u})=\begin{pmatrix}-\Gamma u & -\Gamma\rho \\ -\Lambda & 0 \end{pmatrix}.
\end{equation}
The eigenvalues are given by (in increasing order)
\begin{equation}\label{eq:ActIntEigenvalues}
\lambda_{1/2} = -\Gamma u/2 \mp \sqrt{\Gamma^2 u^2/4 + \Lambda\Gamma\rho},
\end{equation}
and are, of course, functions of $\rho$ and $u$, as are the eigenvectors.
They are also distinct as long as $u$ and $\rho$ are not vanishing simultaneously, so that the matrix is {\it strictly hyperbolic} for $\rho>0$.
Hyperbolicity, which guarantees that the eigenvectors form a basis of the $\mathbf{v}$-space, is a crucial property.
It allows us, by decomposing the initial discontinuity of the Riemann problem in the eigenvector basis, to decompose the problem itself into two simpler ones, in analogy with Eq.~(\ref{eq:LinWaveDecomposition}).

Due to the eigenvectors dependence on $\rho$ and $u$, however, the basis is only local, so that a generic Riemann problem with arbitrarily far left and right states cannot be decomposed along the system eigenvectors.
We will see how to circumvent this problem via the definition of shock and rarefaction curves, which cover the whole $\rho,u$ plane (or, at least, its physical section).
First, we will consider the example of the giant trough initial condition, and learn how to solve the corresponding Riemann problem by combining two shock waves.
Analogously, in the next subsection, we will study rarefaction fans via the giant peak initial condition.
The last subsection deals with the special case of linear degeneracy, occurring when the inclusion density $\rho$ vanishes.

\subsubsection{Shock waves in the giant trough}\label{sssec:shocks}

\begin{figure}[t!]
	\centering
	\includegraphics[angle=-90,width=1.\textwidth]{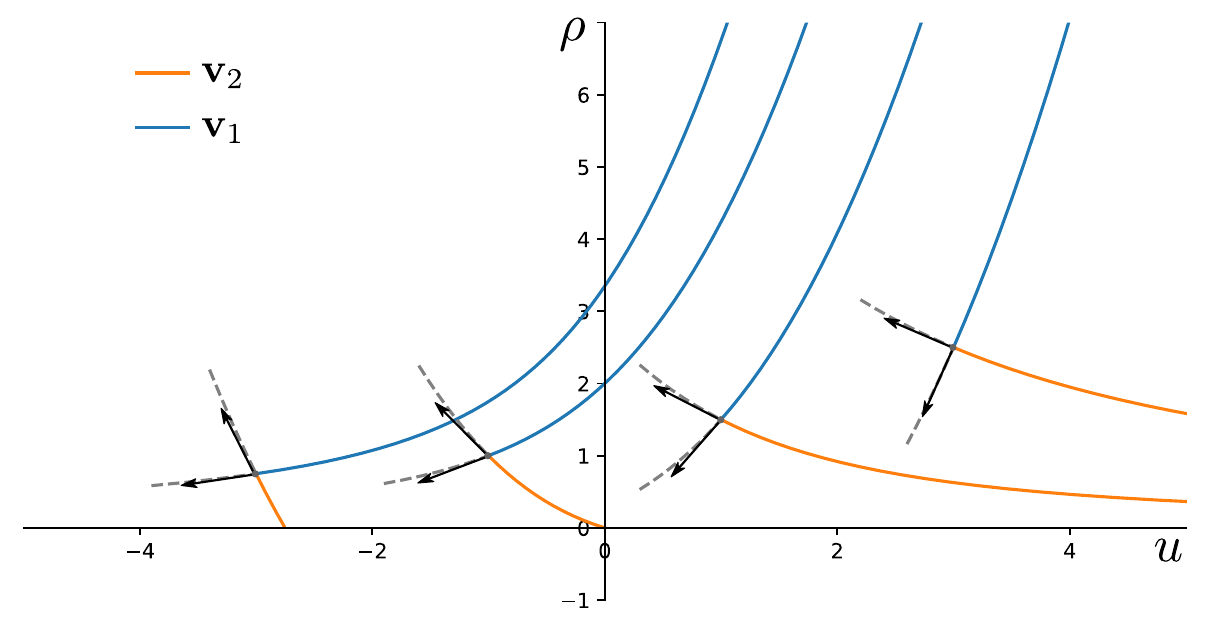}
	\caption{\footnotesize{Shock curves of the active interface equations for several left states $\mathbf{v}_l$ (see discussion in the text). $\mathbf{v}^s_1$ ($\mathbf{v}^s_2$) denotes the graph of the function $\rho_1(u)$ ($\rho_2(u)$), i.e the first (second) shock curve. The grey dashed lines mark the portion of the curves discarded due to not satisfying the Lax entropy condition. The arrows represent the $\mathbf{A}$ matrix right eigenvectors, and point towards the direction of increasing eigenvalue.}}
	\label{fig:ActIntShock}
\end{figure}

The discontinuous initial condition of the Riemann problem can evolve into a either rarefaction fan or a shock wave, or possibly into a superposition of the two.
The Rankine-Hugoniot condition Eq.~(\ref{eq:1DRankineHugoniot}) for the shock speed carries over to the multi-dimensional case, where it reads
\begin{equation}\label{eq:nDRankineHugoniot}
 \mathbf{J}(\mathbf{v}_r)-\mathbf{J}(\mathbf{v}_l)=\sigma\left(\mathbf{v}_r-\mathbf{v}_l\right).
\end{equation}
The perspective taken in the multidimensional Riemann problem is, however, opposite to that adopted in the scalar case.
Rather than computing the speed {\it a posteriori} given $\mathbf{v}_l$ and $\mathbf{v}_r$, we shall fix $\mathbf{v}_l$ and use Eq.~(\ref{eq:nDRankineHugoniot}) as the definition of a curve in the $\mathbf{v}$-plane.
By plugging $J_1=-\Gamma\rho u$ ($\rho$ component) and $J_2=-\Lambda\rho$ ($u$ component, cf. Eq.~\ref{eq:InviscidActInt}) into Eq.~(\ref{eq:nDRankineHugoniot}) one gets
\begin{equation*}
 \begin{aligned} \sigma (\rho-\rho_l) &= -\Gamma(\rho u-\rho_lu_l), \\ \sigma(u-u_l)&=-\Lambda(\rho-\rho_r),\end{aligned}
\end{equation*}
where, to stress that only the left state $\mathbf{u}_l=(\rho_l,u_l)$ is fixed, we have omitted the subscript from the right state.
Once $\sigma$ is eliminated from the equations, there remains a quadratic equation for $\rho$ as a function of $u$ (or viceversa) and $\mathbf{u}_l$.
The two solutions for $\rho$ are
\begin{equation}\label{eq:ActIntShock}
\begin{aligned}
 \rho_1(u;\mathbf{v}_l) = \rho_l + \frac{u-u_l}{\Lambda}\left[ \Gamma u/2 + \sqrt{\Gamma^2u^2/4 + \Gamma\Lambda\rho_l} \right],\\
 \rho_2(u;\mathbf{v}_l) = \rho_l + \frac{u-u_l}{\Lambda}\left[ \Gamma u/2 - \sqrt{\Gamma^2u^2/4 + \Gamma\Lambda\rho_l} \right].
\end{aligned}
\end{equation}
The graphs of $\rho_1(u)$ and $\rho_2(u)$, shown in Fig.~\ref{fig:ActIntShock}, are the aforementioned curves---they are called {\it shock curves}.
Let us call them $\mathbf{v}^s_{1/2}$ and parametrise them with $u$ e.g., $\mathbf{v}^s_{1}(u;\mathbf{v}_l)$ has components $(\rho_1(u;\mathbf{v}_l),u)$.

By construction, the $(\mathbf{v}_l,\mathbf{v}^s_i(u,\mathbf{v}_l))$ Riemann problem is solved by a propagating discontinuity, or a shock.
The shock speed $\sigma_{i}(u;\mathbf{v}_l)$ comes from the Rankine-Hugoniot condition, i.e.
\begin{equation}\label{eq:ActIntShockSpd}
\begin{aligned}
 \sigma_1(u;\mathbf{v}_l)=&-\Gamma u/2 -\sqrt{\Gamma^2 u^2/4 + \Lambda\Gamma\rho_l},\\
 \sigma_2(u;\mathbf{v}_l)=&-\Gamma u/2 +\sqrt{\Gamma^2 u^2/4 + \Lambda\Gamma\rho_l}.
\end{aligned}
\end{equation}
By performing a $\rho,u\rightarrow \rho_l,u_l$ limit of Eq.~(\ref{eq:nDRankineHugoniot}--\ref{eq:ActIntShockSpd}), it can be shown that the $i$-th shock curve tangent tends to the $i$-th eigenvector of $\mathbf{A}(\mathbf{v}_l)$ (see Fig.~\ref{fig:ActIntShock}), whereas $\sigma_i$ tends to the corresponding eigenvalue.
Since the eigenvalues play the role of characteristic slopes, it is  natural to extend the Lax condition Eq.~(\ref{eq:1DLaxEntropy}) as~\cite{lax1973hyperbolic}
\begin{equation}\label{eq:nDLaxEntropy}
  \lambda_i(\mathbf{v}_l) \geq \sigma_i(u;\mathbf{v}_l) \geq \lambda_i(\mathbf{v}^s_i(u,\mathbf{v_l})).
\end{equation}
In order to meet these Lax conditions  with shock speeds  Eq.~(\ref{eq:ActIntShockSpd}), the portion of the shock curves with $u<u_l$ must be discarded.
In other words, a shock develops only if the interface slope is higher on the right than on the left.

\begin{figure}[t!]
	\centering
	\includegraphics[angle=-90,width=1.\textwidth]{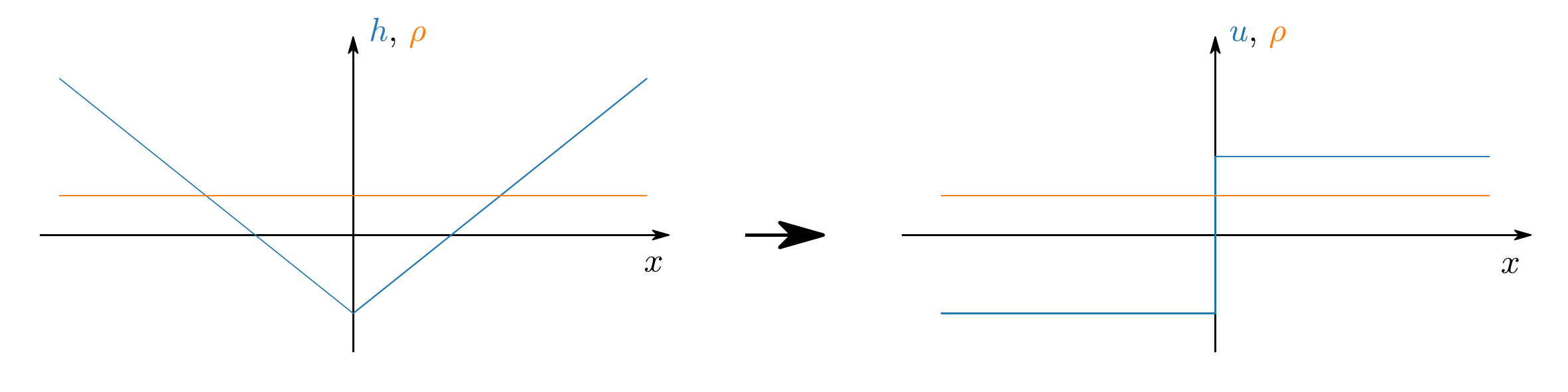}
	\caption{\footnotesize{The giant trough initial condition in the height variable (left panel) becomes a step in the slope variable (right panel).}}
	\label{fig:TroughIC}
\end{figure}

\begin{figure}[t!]
	\centering
	\includegraphics[width=.45\textwidth]{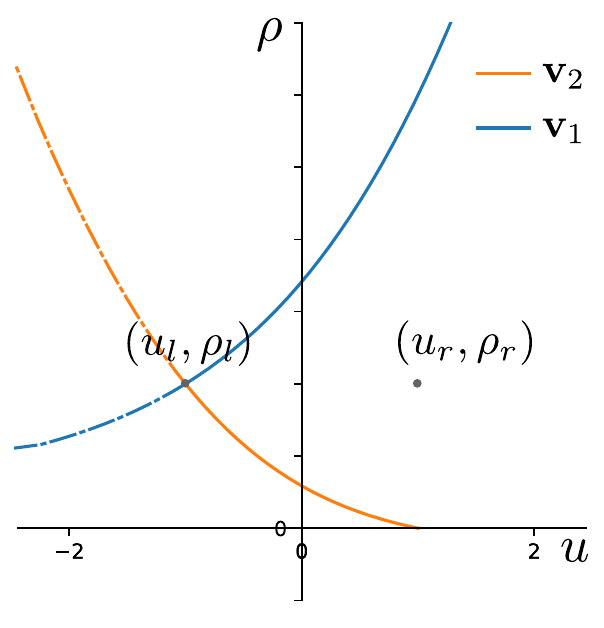}
	\includegraphics[width=.45\textwidth]{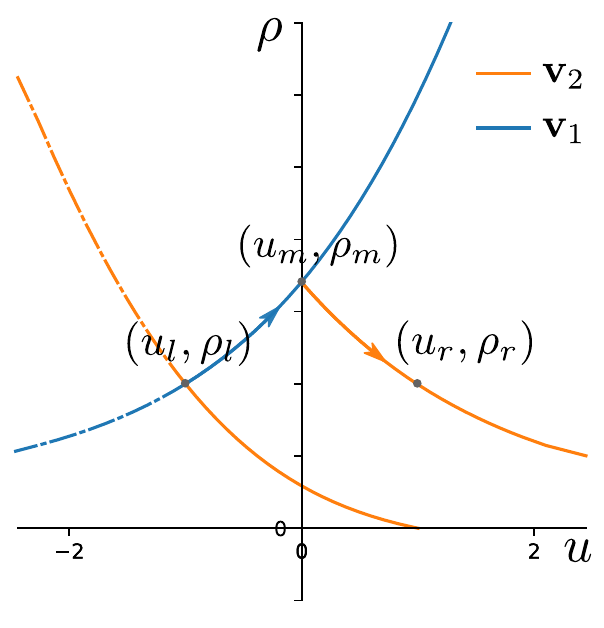}
	\caption{\footnotesize{Schematic representation of the giant trough Riemann problem in the $\mathbf{v}$-plane. The relevant part of the shock curves are shown as solid lines. The intermediate state must be reached via the curve $\mathbf{v}^s_1$, as shown in the right panel, so that the two waves the full solution will consist of do not collide.}}
	\label{fig:TroughICshock}
\end{figure}

We are now able to compute the evolution of the giant trough initial condition (see Fig.~\ref{fig:TroughIC}), which we already used for the linearised equations, for the inviscid active interface equations.
The solution amounts to combining two shock waves travelling in opposite directions.  
In the Riemann problem language, the initial condition is $u_l=-1, \rho_l=\rho_0$ and $\rho_r=\rho_l=\rho_0$ but $u_r=+1$, so that $u_r>u_l$.
As shown in Fig.~\ref{fig:TroughICshock} left panel, $\mathbf{v}_r$ is neither on $\mathbf{v}^s_1$ nor $\mathbf{v}^s_2$.
We will then proceed  by decomposing $\mathbf{v}_r-\mathbf{v}_l$ along the system right eigenvectors.
The only difference with the linear case is that the eigenvectors depend on $\mathbf{v}$, thus we will not connect $\mathbf{v}_r$ to $\mathbf{v}_l$ with two straight lines but with two curves---the shock curves.
Specifically, we will move along $\mathbf{v}^s_1(u;\mathbf{v_l})$, until we hit the point at which $\mathbf{v}_m=(\rho_1(u_m;\mathbf{v}_l),u_m)$ such that $\mathbf{v}_r$ lie  on the second shock curve emanating from $\mathbf{v}_m$, i.e. $\mathbf{v}^s_2(u_r;\mathbf{v}_m)=\mathbf{v}_r$ (see Fig.~\ref{fig:TroughICshock} right panel).
The equation for  the intermediate state $\mathbf{v}_m$, is
\begin{equation*}
 \mathbf{v}^s_2(u_r;\mathbf{v}^s_1(u_m;\mathbf{v}_l))=\mathbf{v}_r \Rightarrow \rho_2(u_r;\mathbf{v}_m) = \rho_r
\end{equation*}
and it is solved by 
\begin{equation*}
u_m=0,\quad \rho_m=\rho_1(0;\mathbf{v}_l)=\rho_0 +\sqrt{\rho_0\frac{\Gamma}{\Lambda}}.
\end{equation*}
The giant trough Riemann problem solution follows by gluing the two shock waves together,
\begin{equation}\label{eq:TroughSolution}
 \left(\rho,u\right)(x,t) = \left\lbrace\begin{aligned}&\left(\rho_0,-1\right) ,& &x/t\; <-\sqrt{\Lambda\Gamma\rho_0},\\&\left(\rho_0+\sqrt{\frac{\Gamma}{\Lambda}\rho_0},0\right) ,&-\sqrt{\Lambda\Gamma\rho_0}\quad <\; &x/t\; < \sqrt{\Lambda\Gamma\rho_0},\\&\left(\rho_0,+1\right) ,&\sqrt{\Lambda\Gamma\rho_0}
\quad <\; &x/t,\end{aligned}\right.
\end{equation}
where $-$ and $+\sqrt{\Lambda\Gamma\rho_0}$ come from $\sigma_1(u_m;\mathbf{v}_l)$ and $\sigma_2(u_r;\mathbf{v}_m)$, respectively.
Using the first shock curve ($\mathbf{v}_1^s$) to find the intermediate state and the second ($\mathbf{v}_2^s$) to reach $\mathbf{v}_r$ from there guarantees that, as $\sigma_1<\sigma_2$, the two shock waves do not collide.

\begin{figure}[t!]
	\centering
	\includegraphics[angle=-90,width=1.\textwidth]{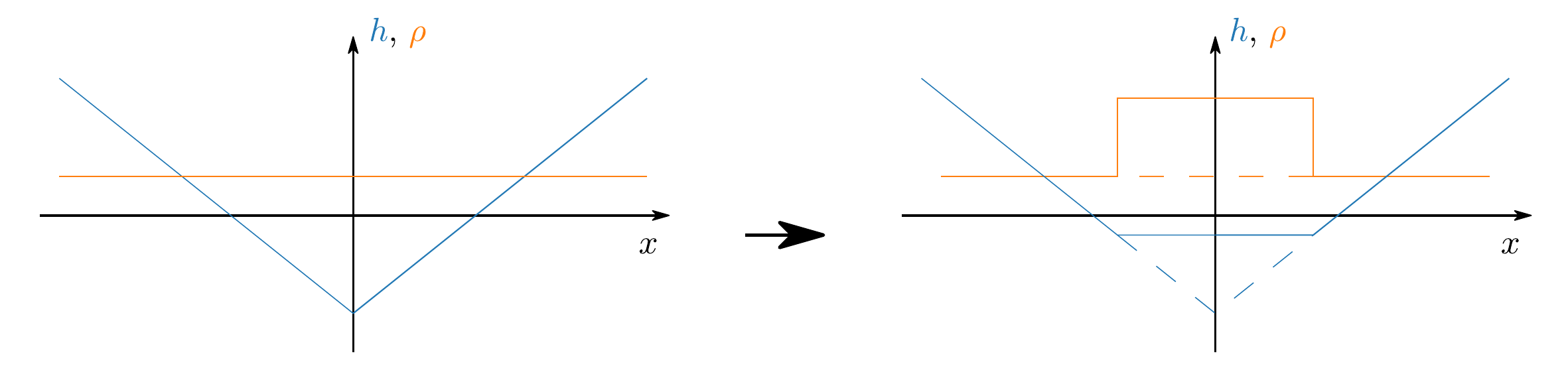}
	\caption{\footnotesize{Evolution of the giant trough initial condition in the height variable. 
The density profile is a top hat function which extends with the shock speeds.
The increased density signals an accumulation of inclusions at the center of the valley, which, consequently, gets filled.  }}
	\label{fig:TroughSolution}
\end{figure}
The solution for the giant trough problem is shown in the height variable in Fig.~\ref{fig:TroughSolution}.
To summarise, two shock waves emanate from the apex of the trough. 
These shock waves delineate a region in which the  inclusion density has a 
top hat profile, indicating an excess inclusion density at the bottom of the trough.
The excess density fills the trough by increasing the height between the shock fronts.
Having learnt how an initial trough generates an inclusion cluster that eventually fills it, it is natural to ask what happens if the initial trough is replaced by a peak.

\subsubsection{Rarefaction fans on the giant peak}\label{sssec:raref}

When $u_r<u_l$ the Lax condition is not satisfied.
As for the single conservation law, a step-like initial condition will smooth out as a rarefaction fan, rather than propagating as a shock wave.
This can be checked by plugging into Eq.~(\ref{eq:InviscidActInt}) the functional form of a rarefaction fan $\mathbf{v}(x,t)=\mathbf{w}(x/t)$.
The equation becomes
\begin{equation}\label{eq:RarefactionFan1}
 \left[\mathbf{A}(\mathbf{w}(\xi))-\xi\mathbf{I}\right]\cdot\mathbf{w}'(\xi)=0,
\end{equation}
in the variable $\xi=x/t$ (the prime denotes derivative w.r.t. $\xi$).
According to Eq.~(\ref{eq:RarefactionFan1}), $\mathbf{w}'$ is the right eigenvector of $\mathbf{A}(\mathbf{w})$ associated with the eigenvalue $\xi$, or, compactly,
\begin{equation}\label{eq:RarefactionFan2}
 \left\lbrace \begin{aligned} &\mathbf{v}'(\xi)=\mathbf{r}_i(\mathbf{v}(\xi)),\\ &\lambda_i(\mathbf{v}(\xi))=\xi.\end{aligned}\right.
\end{equation}
$\lambda_i(\mathbf{w}(\xi))=\xi$ is, in fact, a condition on the $i$th eigenvector normalisation.
Differentiation of both sides w.r.t. $\xi$ yields $(\nabla_{\mathbf{v}}\lambda_i)\cdot\mathbf{r}_i=1$.
The latter condition can be met by appropriate normalisation of the eigenvectors, provided
\begin{equation}\label{eq:GenuineNonLinearity}
 (\nabla_{\mathbf{v}}\lambda_i)\cdot\mathbf{r}_i \neq 0.
\end{equation}
The above inequality is called the {\it genuine nonlinearity condition}, and we will assume it to hold.
Violations will be considered in subsection~\ref{sssec:lindegen}.

\begin{figure}[t!]
	\centering
	\includegraphics[angle=-90,width=1.\textwidth]{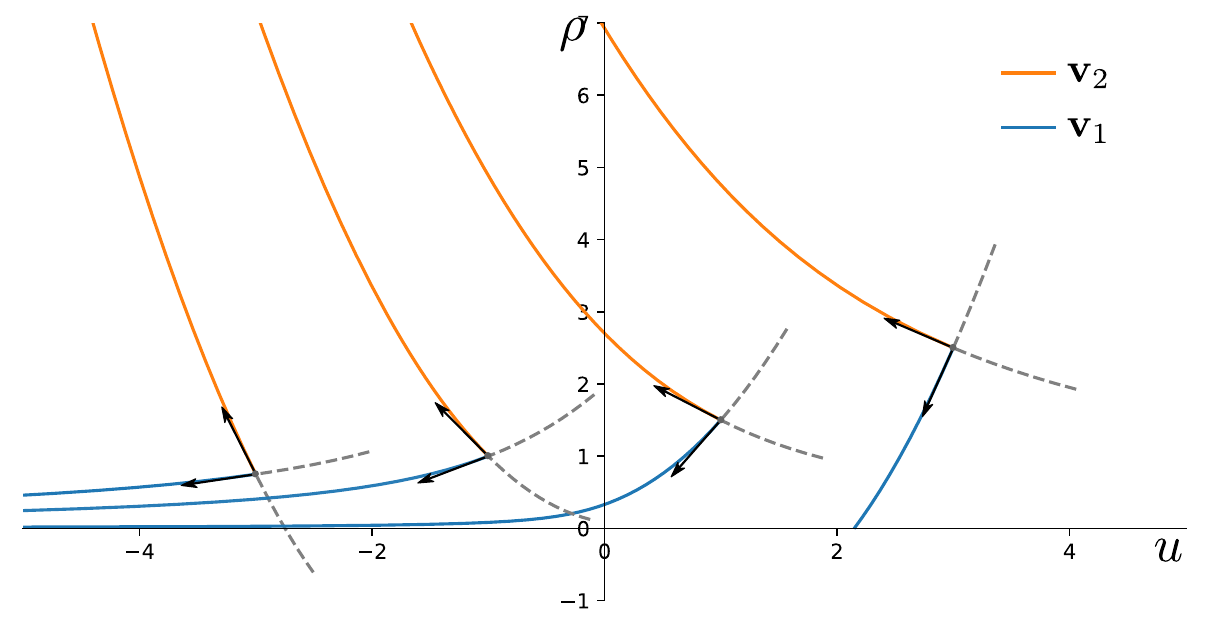}
	\caption{\footnotesize{Rarefaction curves of the active interface equations for several left states $\mathbf{v}_l$ (see discussion in the text). $\mathbf{v}^r_1$ ($\mathbf{v}^r_2$) denotes the first (second) rarefaction curve. The grey dashed lines mark the portion of the curves discarded due to the increasing eigenvalue constraint. The arrows represent the $\mathbf{A}$ matrix right eigenvectors, and point towards the direction of increasing eigenvalue.}}
	\label{fig:ActIntRarefaction}
\end{figure}
The normalised right eigenvectors of the matrix (Eq.~\ref{eq:ActIntMatrix}) are 
\begin{equation}\label{eq:ActIntNormREigen}
 \mathbf{r}_1=\begin{pmatrix} -\frac{\sqrt{\Gamma^2 u^2/4 + \Lambda\Gamma\rho}}{\Lambda\Gamma}\\[1ex] \frac{\sqrt{\Gamma^2 u^2/4 + \Lambda\Gamma\rho}}{\Gamma\lambda_1} \end{pmatrix},\quad \mathbf{r}_2=\begin{pmatrix} \frac{\sqrt{\Gamma^2 u^2/4 + \Lambda\Gamma\rho}}{\Lambda\Gamma}\\[1ex] -\frac{\sqrt{\Gamma^2 u^2/4 + \Lambda\Gamma\rho}}{\Gamma\lambda_2} \end{pmatrix}.
\end{equation}
Each will give rise to an equation such as Eq.~(\ref{eq:RarefactionFan2}).
In order to find the appropriate boundary conditions, we resort to the same approach as the previous subsection: let us fix $\mathbf{v}_l$ (the left initial vector of the Riemann problem) and use Eq.~(\ref{eq:RarefactionFan2}) to find two more curves in the $\mathbf{v}$-plane which are the  {\it rarefaction curves}.
As with the shock curves, we desire  that they  emanate from $\mathbf{v}_l$, hence we will use the latter as initial condition of Eq.~(\ref{eq:RarefactionFan2}).
The two solutions---let us call them $\mathbf{v}^r_1$ and $\mathbf{v}^r_2$---are
\begin{equation}\label{eq:ActIntRarefaction1}
 \begin{pmatrix} \rho^r_1(\xi;\mathbf{v}_l) \\[1ex] u^r_1(\xi;\mathbf{v}_l)  \end{pmatrix} = \begin{pmatrix} \frac{\xi^2}{3\Gamma\Lambda} - \frac{\sqrt{\xi\xi_{1,l}}}{\Lambda}\left[u_l+\frac{2\xi_{1,l}}{3\Gamma}\right] \\[1ex] \sqrt{\frac{\xi_{1,l}}{\xi}}\left[u_l+\frac{2\xi_{1,l}}{3\Gamma}\right] -\frac{2\xi}{3\Gamma} \end{pmatrix},
\end{equation}
and
\begin{equation}\label{eq:ActIntRarefaction2}
 \begin{pmatrix} \rho^r_2(\xi;\mathbf{v}_l) \\[1ex] u^r_2(\xi;\mathbf{v}_l)  \end{pmatrix} = \begin{pmatrix} \frac{\xi^2}{3\Gamma\Lambda} + \frac{\sqrt{\xi_{2,l}\xi}}{\Lambda}\left[u_l+\frac{2\xi_{2,l}}{3\Gamma}\right] \\[1ex] \sqrt{\frac{\xi_{2,l}}{\xi}}\left[u_l+\frac{2\xi_{2,l}}{3\Gamma}\right] -\frac{2\xi}{3\Gamma} \end{pmatrix}.
\end{equation}
For both $i=1$ and $2$, $\xi_{i,l}$ is such that $\mathbf{v}^r_i(\xi_{i,l};\mathbf{v}_l)=\mathbf{v}_l$.
In agreement with $(\nabla_{\mathbf{v}}\lambda_i)\cdot\mathbf{r}_i=1$, only the portion of the rarefaction curve along which the corresponding eigenvalue increases shall be retained.
As, always due to $(\nabla_{\mathbf{v}}\lambda_i)\cdot\mathbf{r}_i=1$, $\xi_{i,l}=\lambda_i(\mathbf{v}_l)$, the eigenvalue increases in the direction of increasing $\xi$.
Hence $u^r_i \leq u_l$, i.e. the rarefaction curves extend in the direction of decreasing $u$ or, in other terms, an initial discontinuity consisting of a drop in the value of $u$ will give rise to a rarefaction fan.

\begin{figure}[t!]
	\centering
	\includegraphics[width=.45\textwidth]{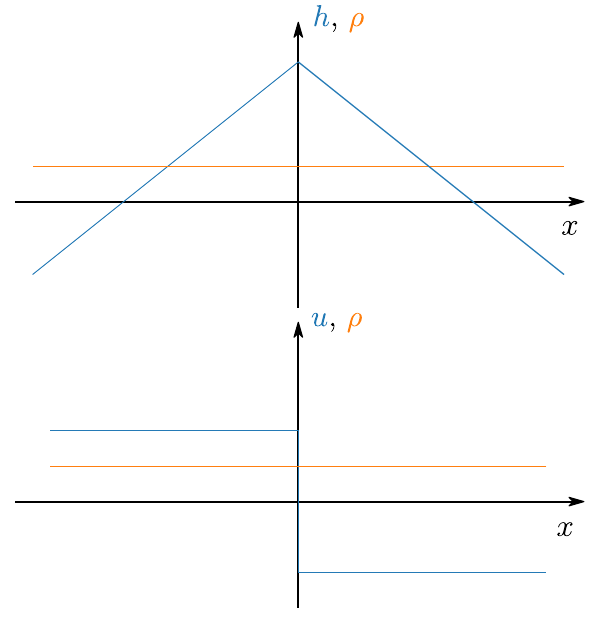}
	\includegraphics[width=.45\textwidth]{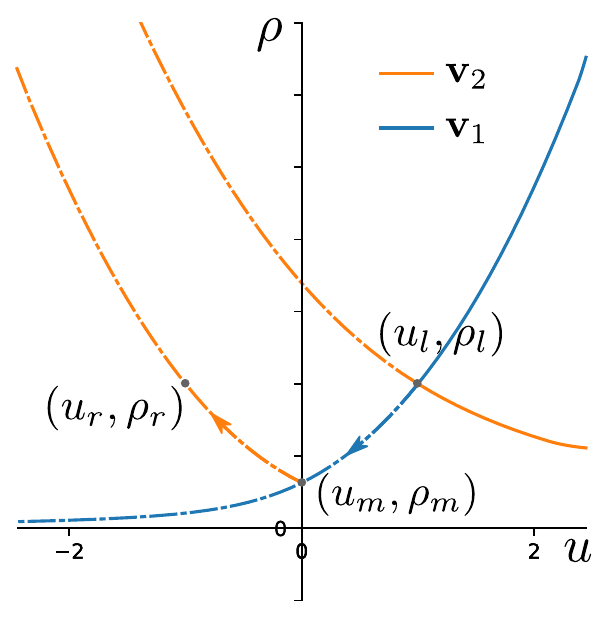}
	\caption{\footnotesize{Giant peak initial condition in the slope and height variable (on the left) and corresponding construction of the solution in the $\mathbf{v}$-plane (on the right). With respect to the giant trough of Fig. \ref{fig:TroughSolution}, the solution will be a double rarefaction fan, with an intermediate density lower rather than higher than the initial one.}}
	\label{fig:HillICRaref}
\end{figure}
The active interface equations rarefaction curves are shown in Fig.~\ref{fig:ActIntRarefaction}.
By construction, the $(\mathbf{v}_l,\mathbf{v}^r_i(\xi,\mathbf{v}_l))$ Riemann problem is solved by a rarefaction fan, whose shape also depends on $\mathbf{v}^r_i(\xi;\mathbf{v}_l)$:
\begin{equation*}
 \mathbf{v}(x,t) = \left\lbrace \begin{aligned} &\mathbf{v}_l, &x/t < \xi_{i,l}\equiv\lambda_i(\mathbf{u}_l)\\ &\mathbf{v}^r_i(x/t), &x/t \in [\xi_{i,l},\xi]\equiv[\lambda_i(\mathbf{v}_l),\xi]\\&\mathbf{v}^r_i(\xi;\mathbf{v}_l), &x/t > \xi.\end{aligned}\right.
\end{equation*}

We are now in a position to solve the giant peak Riemann problem $u_l=1$, $u_r=1$, $\rho_r=\rho_l-\rho_0$ (see Fig.~\ref{fig:HillICRaref}) for the inviscid active interface equations.
The solution amounts to combining two rarefaction waves travelling in opposite directions.  
The procedure is analogous to that used for shock waves, hence we will not explain it in detail.
The solution reads
\begin{equation}\label{eq:PeakICSolution}
 \left(\rho,u\right)(x,t) = \left\lbrace\begin{aligned}&\left(\rho_0,+1\right) ,& &x/t<\lambda_1(\rho_0,+1),\\&\left(\rho^r_1(x/t),u^r_1(x/t)\right) ,&\lambda_1(\rho_0,+1)<&x/t<\lambda_1(\rho_1(\bar{\xi}),0),\\&\left(\rho_1(\bar{\xi}),0\right) ,&\lambda_1(\rho^r_1(\bar{\xi}),0)<&x/t<\lambda_2(\rho^r_1(\bar{\xi}),0),\\&\left(\rho^r_2(x/t),u^r_2(x/t)\right) ,&\lambda_2(\rho^r_1(\bar{\xi}),0)<&x/t<\lambda_2(\rho_0,-1),\\&\left(\rho_0,-1\right) ,&\lambda_2(\rho_0,-1)<&x/t,\end{aligned}\right.
\end{equation}
where $\bar{\xi}$ is such that $\mathbf{v}_1^r(\bar{\xi})=\mathbf{v}_m$ and is found by solving
\begin{equation*}
 \mathbf{v}_r = \mathbf{v}^r_2(\lambda_2(\rho_r,u_r);\mathbf{v}^r_1(\bar{\xi};\mathbf{v}_l)).
\end{equation*}

The solution for the giant peak initial condition is shown in Fig.~(\ref{fig:PeakSolution}), left panel: it consists of two rarefaction waves emanating from the apex of the peak.
In propagating, these rarefaction waves leave behind a region where the density is reduced (see the flat-bottomed trough profile in the left panel of the figure) and the height profile is smoothened (the flat solid line in the figure replaces the sharp, dashed wedge).
There is, however, a complication that arises when the inclusion density at the bottom of the trough is reduced to zero (Fig.~(\ref{fig:PeakSolution}) right panel).
The density must be physically greater than zero: mathematically, we require $\rho^r_1$ to be positive to satisfy the genuine nonlinearity condition Eq~(\ref{eq:GenuineNonLinearity}).
Due to the functional form of $\rho_1^r(\xi;\mathbf{u}_l)$, the condition translates into
\begin{equation*}
 u_l+\frac{2\lambda_1(\rho_l,u_l)}{3\Gamma} <0 \Leftrightarrow \rho_l > \rho_c \equiv \frac{3\Gamma}{4\Lambda}u_l^2.
\end{equation*}
As long as $\rho_l>\rho_c$ the giant peak Riemann problem is solved as above by Eq.~(\ref{eq:PeakICSolution}).
However, the case $\rho_l \leq \rho_c$ requires the additional concept of {\it linear degeneracy}, which we consider in the next subsection.

\subsubsection{Linear degeneracy at vanishing density}\label{sssec:lindegen}

\begin{figure}[t!]
	\centering
	\includegraphics[angle=-90,width=1.\textwidth]{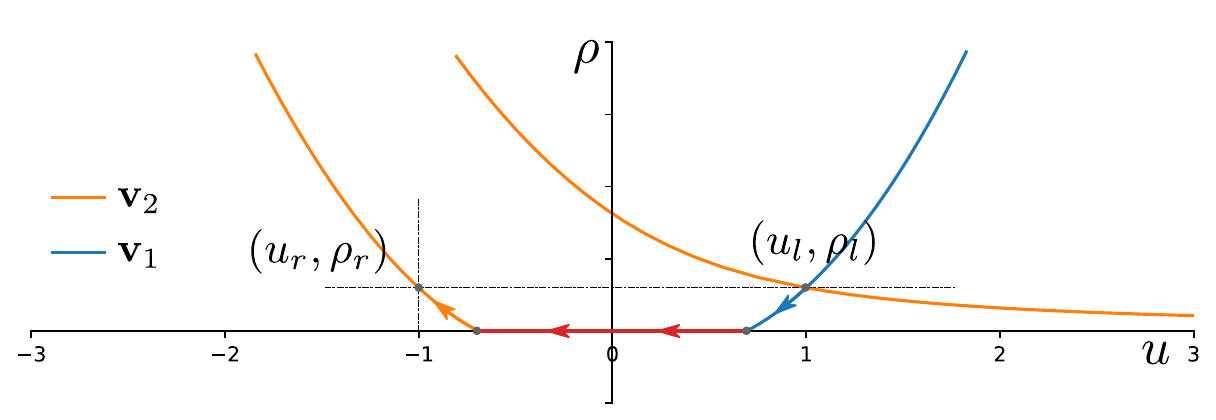}
	\caption{\footnotesize{Schematic construction of the solution shown in Eq.~(\ref{eq:PeakICSolutionBis}). On the $\rho=0$ line the system becomes linearly degenerate. As a result, the shock curve collapse onto a horizontal, straight line and can be extended for $u$ smaller than the starting point (see discussion in the text). To reach the right state $\mathbf{v}_r$, then, one can use the red horizontal line shown in the figure to move along the $u$-axis, down to that point that can be connected directly to $\mathbf{v}_r$ via a rarefaction curve.}}
	\label{fig:ActIntRarefContactDics}
\end{figure}
\begin{figure}[t!]
	\centering
	\includegraphics[angle=-90,width=1.\textwidth]{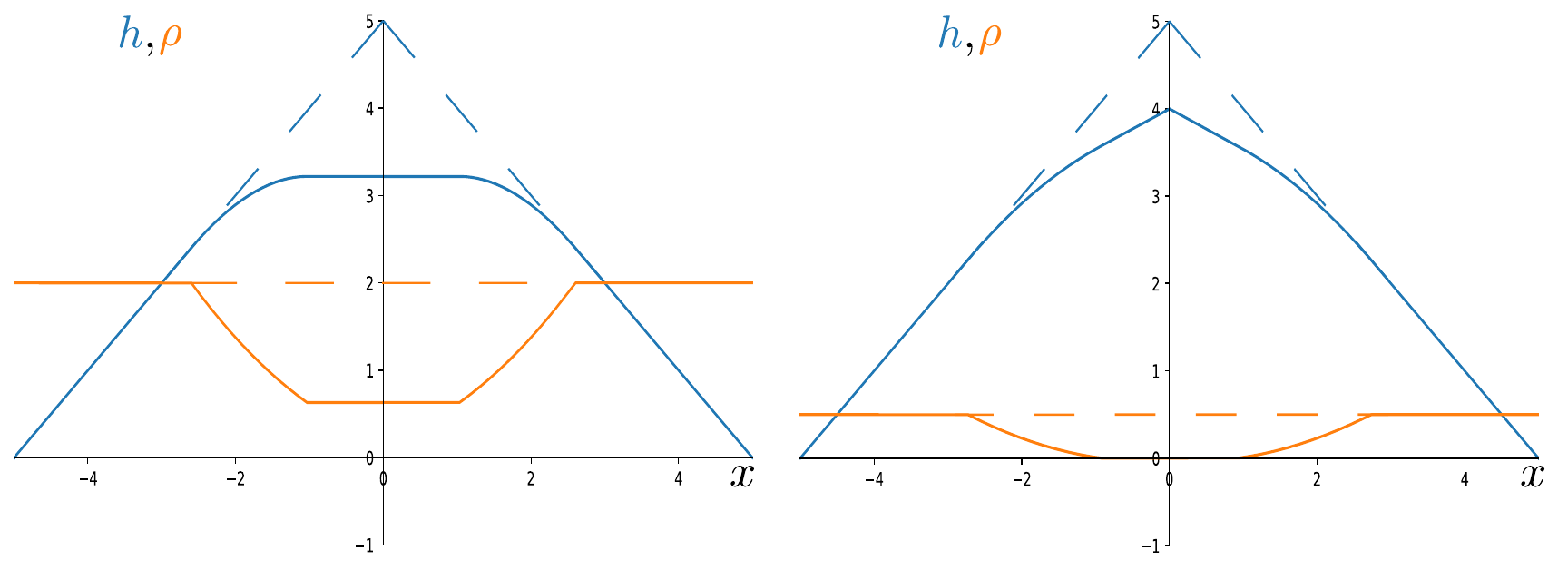}
	\caption{\footnotesize{Solution of the giant peak Riemann problem in the height-density variables (dashed lines represent the initial condition). $\rho_0>\rho_c$ in the left panel (Eq.~(\ref{eq:PeakICSolution})), but not in the right one (Eq.~(\ref{eq:PeakICSolutionBis})). In the right panel, the initial inclusion density is so low that the advection due to the slope completely depletes the peak. Once left with no inclusions, the smoothening of the peak ends.}}
	\label{fig:PeakSolution}
\end{figure}

Consider the giant peak initial condition with $\rho_l<\rho_c$ (see Figure \ref{fig:PeakSolution}).
When leaving $\mathbf{v}_l$ along the first rarefaction curve, there is a $\tilde{\xi}$ such that $\rho_1^r(\tilde{\xi};\mathbf{v}_l)=0$ and $u_1^r(\tilde{\xi};\mathbf{v}_l)=\tilde{u}>0$ (cf. Fig.~\ref{fig:ActIntRarefContactDics}).
As $\rho=0$, both $\lambda_2$ and $(\nabla_{\mathbf{v}}\lambda_2)\cdot\mathbf{r}_2$ vanish.
The  genuine nonlinearity (\ref{eq:GenuineNonLinearity}), then, ceases to hold, and the corresponding rarefaction curve ($\mathbf{v}_2^r$ in this case) is not  defined anymore.
In such cases the pair $\lambda_2,\mathbf{r}_2$ is said to be linearly degenerate.
The reason is that  $\lambda_2$ is constant along the direction of $\mathbf{r}_2$, so that along this direction the conservation laws are effectively linear.
The solution of the problem  then reduces to  a simple transport wave.

This can be shown by considering the second shock curve.  When emanating from a point on the positive $u$ axis
this curve coincides with the horizontal line $\rho=0$ (cf. $\rho_2(u;\mathbf{v}_l)|_{\rho_l=0}$ from Eq.~(\ref{eq:ActIntShock})), and the shock speed $\sigma_2(u,\mathbf{v}_l)$ vanishes. As the second eigenvalue $\lambda_2$ vanish too, the Lax condition Eq.~(\ref{eq:nDLaxEntropy}) is identically satisfied.
Thus, the transport wave with vanishing speed is the physical solution.
For $u<0$ the pair $\lambda_2,\mathbf{r}_2$ meets again the genuine nonlinearity condition, but the pair $\lambda_1,\mathbf{r}_1$ does not.
By repeating the argument used for positive $u$, $u=0$ can be connected with $u=-\tilde{u}$ via the first shock curve $\mathbf{v}_1$ emanating from the $\mathbf{v}$-plane origin (see again Fig.~(\ref{fig:ActIntRarefContactDics})).
As $\sigma_1(u;\mathbf{v}_l)$ vanishes on the half-line $\rho=0,u<0$, there is no inconsistency in moving first along the second shock curve and along the first shock curve later---they are both associated with a static discontinuity.
A  static discontinuity between $u=\tilde{u},\rho=0$ and $u=-\tilde{u},\rho=0$ is also physically reasonable, as the interface is static where there are no inclusions.

Therefore, the solution for the case  $\rho_0 < \rho_c$, where the Riemann problem generates a vanishing inclusion density,  is given by
\begin{equation}\label{eq:PeakICSolutionBis}
 \left(\rho,u\right)(x,t) = \left\lbrace\begin{aligned}&\left(\rho_0,+1\right) ,& &x/t<\lambda_1(\rho_0,+1),\\&\left(\rho_1(x/t),u_1(x/t)\right) ,&\lambda_1(\rho_0,+1)<&x/t<\lambda_1(0,u_1(\hat{\xi})),\\&\left(0,u_1(\hat{\xi})\right) ,&\lambda_1(0,u_1(\hat{\xi}))<&x/t<0,\\&\left(0,-u_1(\hat{\xi})\right) ,&0<&x/t<\lambda_2(0,-u_1(\hat{\xi})),\\&\left(\rho_2(x/t),u_2(x/t)\right) ,&\lambda_2(0,-u_1(\hat{\xi}))<&x/t<\lambda_2(\rho_0,-1),\\&\left(\rho_0,-1\right) ,&\lambda_2(\rho_0,-1)<&x/t.\end{aligned}\right.
\end{equation}
Having studied the two model Riemann problems of the giant trough and the giant peak, we are finally in the position of enforcing periodic boundary conditions on the inviscid active interface equations, and build a solution that takes the system finiteness into account.
This will be the topic of section \ref{sec:osc}.

\section{Oscillating dynamics with periodic boundaries}\label{sec:osc}

\begin{figure}[t!]
	\centering
	\includegraphics[angle=-90,width=1.\textwidth]{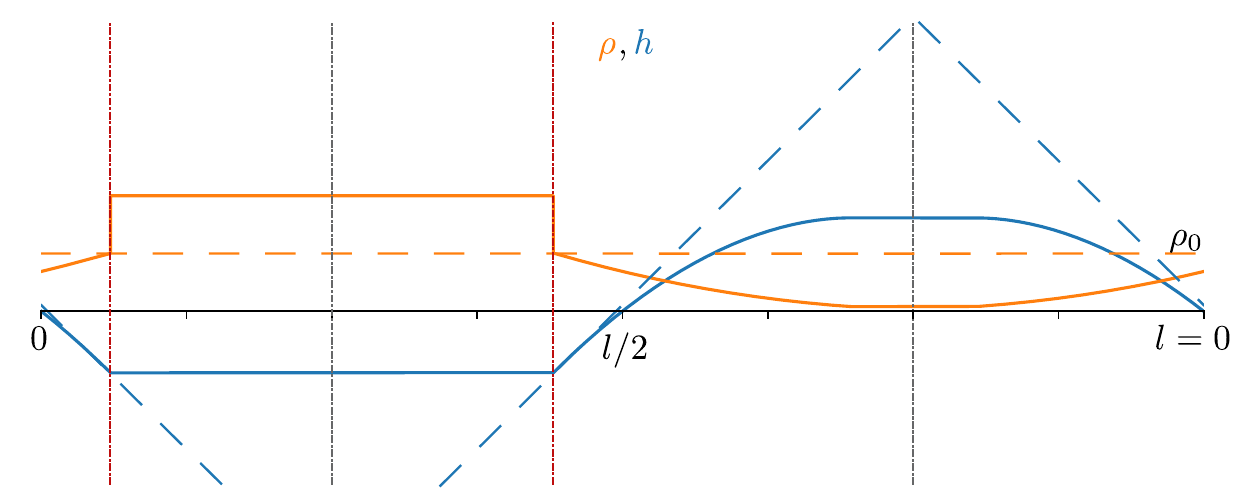}
	\caption{\footnotesize{Interface (blue) and density (orange) profiles at the time of the first waves collision $\tau$. The profiles come from Eq.(\ref{eq:TroughSolution}) and (\ref{eq:PeakICSolution}-\ref{eq:PeakICSolutionBis}). The initial condition is also shown with dashed lines. Two vertical, gray, dot-dashed lines mark the position of the discontinuities in the initial condition, while two red ones highlights the location of the discontinuities at time $\tau$.}}
	\label{fig:Oscillations1}
\end{figure}

\begin{figure}[t!]
	\centering
	\includegraphics[angle=-90,width=1.\textwidth]{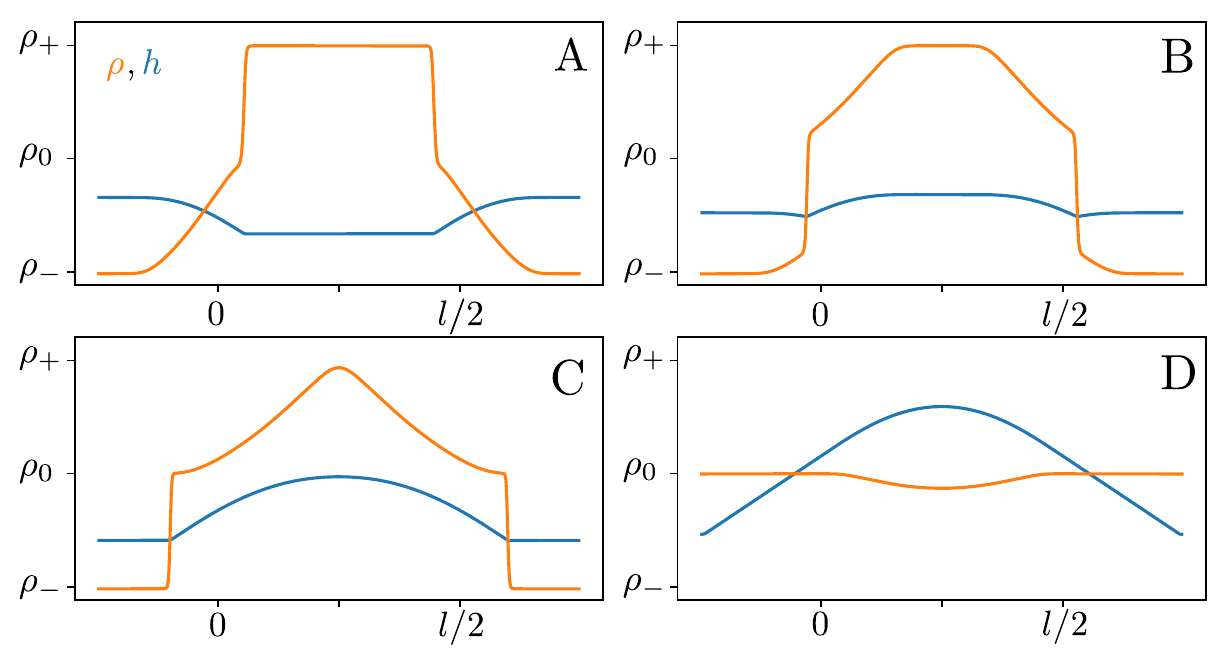}
	\caption{\footnotesize{Snapshot of the density (orange) and height (blue) profiles at several times, all greater than or equal to the time of the first collision $\tau$. Snapshot A is taken exactly at time $\tau$ at which the shocks emanating from thr trough and rarefaction fans emanating from the peak  meet. Snapshot B shows the system state after the shocks have passed through the front of the rarefaction waves: the density in the middle decreases and the new peak starts to form. Snapshot C is taken when the shocks meet the tail of the original rarefaction curves. The dynamics, with respect to shorter times $t<\tau$, is reversed: a new peak, located at $l/4$ is formed with a rarefaction wave, while a shock wave generates a trough at $3l/4$. Lastly, panel D shows a snapshot right before all the waves have travelled the whole system size. The density goes back to uniform, but the peak and the trough of the initial condition have been swapped. From this state, the whole cycle starts again.}}
	\label{fig:Oscillations2}
\end{figure}

In section \ref{sec:Inviscid} we have assumed the system to be infinite.
We now consider periodic boundary conditions with an initial condition  that, due to periodicity, is a combination of the giant peak and the giant trough considered in the previous section (see Fig.~\ref{fig:Oscillations1}).
The  dynamics is most easily visualised by a numerical solution of the inviscid equations.
The solution, obtained by a discretisation of time and space, is shown in Fig.~\ref{fig:Oscillations2} and we shall refer to this in the following discussion.

Initially, the analytical solutions (\ref{eq:TroughSolution}) and (\ref{eq:PeakICSolution}-\ref{eq:PeakICSolutionBis}) are still valid.
The inclusions move away the peak, which smoothens as a rarefaction wave, while accumulating at the trough.
At the same time, the height at the bottom of the trough rises due to the increased inclusion density, while the peak flattens.
The result of this dynamics is depicted in Fig.~\ref{fig:Oscillations1}.
We placed the initial condition's discontinuities at $x=l/4$ and $3l/4$, where $l$ is the system size (see the vertical gray, dot-dashed lines in the figure).
After some time $t$, they are located at (by recalling the shock speeds) $x_{t,1}=l/4 + \sqrt{\Gamma\Lambda\rho_0}t$, $x_{t,2}=l/4 - \sqrt{\Gamma\Lambda\rho_0}t$, i.e. close to $0$ and $l/2$.
The physical interpretation of the top hat density profile is that an inclusion aggregate forms at the trough position and spreads laterally with speed $\pm\sqrt{\Gamma\Lambda\rho_0}$ (cf. Eq.~(\ref{eq:TroughSolution})).
The excess density of the aggregate is $\sqrt{\rho_0\Gamma/\Lambda}$.
As we have shown in~\cite{cagnetta2018aa}, these values fit well with the inclusion clusters size and speed measured in the microscopic, stochastic model.

As the trough triggers the two shock waves, the peak triggers two rarefaction waves, which travel at speed $\pm\left(\Gamma/2+\sqrt{\Gamma^2/4 + \Lambda\Gamma\rho_0}\right)$ towards the shocks fronts.
The meeting of shocks and rarefaction waves occurs at time
\begin{equation}\label{eq:OscPeriod}
 \tau = \frac{l}{2\left(\Gamma/2+\sqrt{\Gamma^2/4 + \Lambda\Gamma\rho_0} + \sqrt{\Lambda\Gamma\rho_0}\right)} .
\end{equation}
The system state at the meeting time is shown in the Fig.~\ref{fig:Oscillations1} and also presented  in panel A of Fig.~\ref{fig:Oscillations2}.
Panel B, instead, shows the height and density profiles right after the meeting time.
Notice how the shock fronts are still propagating away from the high-density region, but now the density at the front is lower than that in the middle.
The profiles connecting the bulk of the high density-region with the shock fronts are remnants of the rarefaction fans, which are now split in half by the shocks.
Notice, also, how the interface is forming a peak in the high-density region, to replace the trough of the initial condition.
Although the new discontinuities (highlighted in the figure by vertical red lines) do not strictly constitute a Riemann problem as the values of $\rho$ and $u$ on their sides are not constant,
they  can nevertheless  give us  some insight on the system behaviour for $t>\tau$.

Specifically, we will proceed as if we were solving two new Riemann problems with
\begin{equation*}
 \mathbf{v}_l =\mathbf{v}(x_{\tau,1}^-,\tau),\quad \mathbf{v}_r =\mathbf{v}(x_{\tau,1}^+,\tau),
\end{equation*}
and the same for the discontinuity at $x_{\tau,2}$.
For the first discontinuity, for instance, the values on the left are $u_l=-1$ and $\rho_l=\rho_0$, as in the giant trough of the last section.
The shock and rarefaction curves emanating from such $\mathbf{v}_l$ are shown in Fig.~\ref{fig:TroughICshock}.
The values on the right are $u=0,\rho=\rho_0+\sqrt{\rho_0\Gamma/\Lambda}$---nothing but the intermediate state $\mathbf{v}_m$ of Fig.~\ref{fig:TroughICshock}, right panel.
The shock is then initially preserved, and also its speed remains the same: $-\sqrt{\Gamma\Lambda\rho_0}$.
As it progresses, however, the left state changes, with $\rho$ decreasing and $u$ shifting towards zero.
With respect to Fig.~\ref{fig:TroughICshock}, the point $\mathbf{v}_l$ moves towards the origin of the axes.
Specifically, it does so by following the orange segment with an arrow of the left panel of Fig.~\ref{fig:HillICRaref}, that is the second rarefaction curve (recall this is how we have determined the rarefaction fan profile).
As soon as $\mathbf{v}_l$ moves down, it cannot be connected to $u=0,\rho=\rho_0+\sqrt{\rho_0\Gamma/\Lambda}$ with a single shock curve anymore.
A rarefaction curve can be used to bridge the gap.
This rarefaction curve needs to be that associated with the second eigenspace, as the shock curve  is that  associated with the first, and it gives rise to the rarefaction fan connecting the high density region in the middle of panel B with the shock front.
Putting the theory aside for a moment, it is as if the shock wave and the front of the rarefaction fan pass through each other. In doing so, the rarefaction fan acquires the discontinuity of the shock, while the shock lowers its speed due to the lower density found after the rarefaction fan.

Thus, after passing through the front of the rarefaction wave,  the shock keeps propagating albeit with a slightly different speed, until it meets the tail of the rarefaction wave, which travels at speed $+\sqrt{\Gamma\Lambda\rho_1^r(\bar{\xi})}$ (cf. Eq.~(\ref{eq:PeakICSolution})).
The snapshot shown in panel C is taken at the time when the shocks meet the tails of the original rarefaction waves.
Here the density depletion has progressed---at the front, particularly, the density is the same as the initial condition---and the height peak at $l/4$ is even more pronounced.
This collision can be understood again in terms of a new Riemann problem, this time with initial discontinuity $\mathbf{v}_l=(0,\rho_1^r(\bar{\xi})<0)$ and $\mathbf{v}_r=(+1,\rho_0)$.
The result is again a combination of a shock and a rarefaction wave, with the shock that will now fill the low-density region while forming a new trough at $3l/4$ (notice the initial condition had a peak here), while the rarefaction wave will keep decreasing the density and building the peak at $l/4$.
Though we described the waves close to $x=0$, the phenomenology is the same at $x=l/2$: it suffices to swap left and right states and change the sign of the wave speeds.

Thus, there are now two shocks travelling towards each other, as are  the two  rarefaction waves.
The two collisions will take place at $x=3l/4$ and $l/4$, for shocks and rarefaction fans, respectively (see panel D of Fig.~\ref{fig:Oscillations2}).
The system now looks like a flipped version of the one we started with, with the density returning to  a uniform value  $\rho_0$, but the interface peak at $l/4$ and the trough at $3l/4$.
Notice, however, the new peak is not as sharp as that of the initial condition and that $\rho$ has not quite attained the value $\rho_0$ everywhere.
The reason is that, in the numerical solution, we have added a small viscous term (as small as the lattice spacing of the spatial discretisation).
This small viscous term will cause dissipation and at every iteration of the dynamics just described the interface is slightly flatter than before---it will, ultimately, be flat.

In fact, this dissipation affects the analytical solutions too, where the vanishingly small viscous terms enter through the Lax condition.
To summarise, the giant trough and giant peak initial conditions, combined on a periodic system, cause an oscillatory behaviour.
In the cycle, illustrated in Fig.~\ref{fig:Oscillations2}, the peak and trough reform at diametrically opposite positions, via a pair of shock waves and rarefaction fans travelling around the system and passing through each other.
Eventually, dissipation kicks in, the waves running through the system decay diffusively, so that the density becomes uniform and the interface flat.
The addition of noise could prevent this  trivial outcome:
by creating random kinks in the interface and displacing the inclusions, it will generate small peaks and troughs and density inhomogeneities that give rise to new waves even after dissipation has completely smoothened the initial condition.

\section{Discussion and conclusions}

In this paper we have solved the inviscid limit of the active interface equations and explained the active interface behaviour at the Euler scale by considering a combination of discontinuous initial conditions.
Extreme interface profiles such as the giant trough and the giant peak considered in sections \ref{sssec:shocks} and \ref{sssec:shocks} are used in kinetic roughening problems to probe the system relaxational dynamics~\cite{kardar1986aa}: we found that the giant trough relaxes via two shock waves emanating from its apex (see Fig. \ref{fig:TroughSolution}), while the peak decays by forming two rarefaction fans (see Fig. \ref{fig:PeakSolution}).
It is worth remarking here that the physical limit of zero density enters the picture as the special mathematical case of linear degeneracy.
Interestingly, while the height profile relaxes the initially homogeneous density profile changes, by developing a top hat profile on the trough and a flat-bottomed trough on the peak, as inclusions move away from the latter to accumulate at the former.
As a result, when the waves originated from the trough and the peak meet, the interface profile is (almost) continuous and flat while the density has a jump discontinuity.

Another suitable Riemann problem shows that this new state also generates two shock waves and two rarefaction fans: ultimately, it is as though the two sets of waves have never met and have simply passed through  each other.
An interesting application of this scenario is to the periodic system considered in section~\ref{sec:osc}.       
As the corresponding shock  and rarefaction waves propagate around the system and pass through each other, the system itself exhibits an oscillatory behaviour illustrated in Fig.~\ref{fig:Oscillations2}.
This, we argue, is the dynamical origin of the oscillations observed in~\cite{cagnetta2018aa}.
Apart from yielding the scaling of the oscillation period, the inclusion aggregate's typical size and the wave speed, our calculations also confirm that the oscillatory behaviour of the active interface is not a simple transient.
On the one hand, as the oscillation period scales with the system size, we can regard the oscillations as the characteristic behaviour at the Euler scale (dynamic exponent $z=1$). 
On the other hand,  due to the small viscous terms in the equations, dissipation eventually dominates the dynamics leading to a stable steady state and  dynamic exponent $z=2$.
Let us remark that this holds true for the simplified active interface equations with no KPZ nonlinearity $\left(\nabla h\right)^2$~\footnote{F. Cagnetta, D. Marenduzzo, M. R. Evans, \textit{Unpublished}.}. If this term is included, KPZ modes should appear~\cite{spohn2015aa} in the height dynamics and possibly influence the inclusion dynamics too.
We also note  that in  other studies of
models comprising two driven conserved densities,  one generally expects the two conserved quantities to be characterised by two dynamical exponents~\cite{spohn2015aa}.

The major question arising from our picture concerns the role of the noise in the system.
As we have already mentioned at the end of section~\ref{sec:osc}, the presence of noise in Eq.~(\ref{eq:ActiveKPZ}) might act as the seed of a non-trivial dynamics, even after the density and interfacial slope profiles have relaxed towards the homogeneous state.
In fact, the oscillatory behaviour we observed in~\cite{cagnetta2018aa} at the microscopic model level was attained by starting from a flat interface and homogeneous density initial condition, and was therefore entirely generated by the noise.
Hence, we speculate that noise is fundamental in keeping the $z=1$ behaviour of the system alive even after the effects of the initial conditions are no longer significant.
It would, of course, be of great interest to study  stochasticity in a systematic way, in order to better
understand  its effects.\\

\noindent  {\bf Acknowledgements} FC acknowledges support from the Scottish Funding Council under a studentship.

\bibliographystyle{unsrt}						
\bibliography{ScalingLimits,Probability,KPZ,PartFlucField,PDE,Paper} 	

\end{document}